\documentclass{emulateapj}\usepackage{timesfonts}

\slugcomment{draft of \today}

\shorttitle{Diffusive Shock Acceleration at SNRs}
\shortauthors{Kang, Jones, and Edmon}

\def\kms{~{\rm km~s^{-1}}}
\def\yrs{~{\rm yrs}}
\def\muG{~{\mu\rm G}}

\begin{document}
\title{Nonthermal Radiation from Supernova Remnants: 
Effects of Magnetic Field Amplification and Particle Escape}

\author{Hyesung Kang$^1$, T. W. Jones$^2$, and Paul P. Edmon$^{3}$}

\affil{$^1$Department of Earth Sciences, Pusan National University, Pusan
609-735, Korea: kang@uju.es.pusan.ac.kr\\
$^2$School of Physics and Astronomy, University of Minnesota, Minneapolis, MN 55455, USA: twj@msi.umn.edu\\ 
$^3$Research Computing, Harvard University, Cambridge, MA 02138, USA: 
pedmon@cfa.harvard.edu
}

\altaffiltext{1}{Author to whom any correspondence should be addressed.}

\begin{abstract}
We explore nonlinear effects of wave-particle interactions on the diffusive shock
acceleration (DSA) process in Type Ia-like, SNR blast waves, 
by implementing phenomenological models for magnetic field amplification,
Alfv\'enic drift, and particle escape in time-dependent numerical simulations of nonlinear DSA.
For typical SNR parameters the CR protons can be accelerated to PeV energies 
only if the region of amplified field ahead of the shock is extensive enough to
contain the diffusion lengths of the particles of interest.
Even with the help of Alfv\'enic drift, 
it remains somewhat challenging to construct a nonlinear DSA model for SNRs 
in which order of 10 \% of
the supernova explosion energy is converted to the CR energy and the magnetic field 
is amplified by a factor of 10 or so in the shock precursor,
while, at the same time, the energy spectrum of PeV protons is steeper than $E^{-2}$.
To explore the influence of these physical effects on observed SNR
emissions, we also compute resulting radio-to-gamma-ray spectra.
Nonthermal emission spectra, especially in X-ray and gamma-ray bands,
depend on the time dependent evolution of CR injection process, 
magnetic field amplification, and particle escape, as well as the shock dynamic
evolution. 
This result comes from the fact that the high energy end of the CR spectrum is composed of the particles that
are injected in the very early stages of blast wave evolution.
Thus it is crucial to understand better the plasma wave-particle interactions associated with 
collisionless shocks in detail modeling of nonthermal radiation from SNRs.

\end{abstract}


\section{INTRODUCTION}

Supernova remnants (SNRs) are strong sources of nonthermal radiations, indicating clearly that they are sites of efficient
particle acceleration. In fact, SNRs are thought to be responsible via  the diffusive shock acceleration (DSA) mechanism for the production of 
most of the Galactic cosmic rays (CRs) at least up to the first knee energy of $10^{15.5}$eV
 \citep[see][for reviews]{hill05, reynolds08}.  The spectral and spatial
distributions of the nonthermal emissions carry important information about how DSA works in SNRs. At present there are several
significant tensions in this comparison, especially in comparisons that account for likely nonlinear feedback of DSA on the shock
dynamics and structure \citep[e.g.][]{malkov11,capri12,kang13}. 
The possibility of strong magnetic field amplification (MFA) as a consequence of nonlinear DSA has recently received
considerable attention in this context \citep[e.g.][]{reynolds12,schure12}. 

In DSA theory suprathermal particles go through pitch-angle scatterings by 
magnetohydrodynamic (MHD) waves around collisionless shocks and can be accelerated to
relativistic energies through the Fermi first-order process \citep{bell78, dru83, maldru01}.
In fact, those waves are known to be self-excited both resonantly and non-resonantly 
by CRs streaming away from the shock \citep[e.g.][]{skill75b, lucek00, bell04}.
Plasma and MHD simulations have shown that the CR streaming instability indeed excites MHD waves 
and amplifies the turbulent magnetic fields by as much as orders of magnitude in the shock precursor 
\citep[e.g.][]{zp08,ohira09,riqu09, riqu10,bell13}.
Thin X-ray rims of several young Galactic SNRs 
provide observational evidence that the magnetic field is amplified up to several $100 \mu$G 
downstream of the forward shock \citep[e.g.][]{bamba03,parizot06,eriksen11}.
We note, however, there is as yet no direct observational evidence for the amplified magnetic field
in the upstream, precursor structures of SNR shocks.

An immediate consequence of magnetic amplification (MFA) in shock precursors is the potential 
to accelerate CR ions beyond the first knee, which is otherwise difficult in SNRs \citep{lag83}.
The maximum attainable energy, given by the so-called Hillas constraint, $E_{\rm max} \sim (u_{\rm s}/c) eZ  B r_{\rm s}$, can reach up 
to $10^{15.5}Z$ eV for typical SNRs only
if the upstream, ISM, magnetic field, $B$, is amplified in the precursor of the shock by a factor of at least 10 or so  above typical ISM
values \citep[e.g.,][]{lucek00, hill05}.
Here $u_{\rm s}$ and $c$ are the shock speed and the light speed, 
respectively, $eZ$ is the particle charge and $r_{\rm s}$ is the shock radius.

Generally this scenario depends on the highest energy CRs themselves driving the amplification of turbulent fields 
over the associated CR diffusion length scale of $l_{\rm max} = \kappa(p_{\rm max})/u_{\rm s} \sim
r_{\rm g}(p_{\rm max}) \cdot (c / 3 u_{\rm s})$, 
which is much larger than the gyroradius, $r_{\rm g}(p_{\rm max}) = p_{\rm max}c/(eZB)$, where $\kappa(p)$ is the CR diffusion coefficient \citep{kang13}.  
The diffusion length, $l_{\rm max}$,  corresponds to the scale height of the shock precursor, which in practical terms, if one adopts Bohm diffusion, is approximately
$l_{\rm max}\approx 0.65{\rm pc}(p_{\rm max}c/1{\rm PeV})(B_1/50\muG)^{-1} ~
(u_{\rm s}/3000\kms)^{-1}$ 
(where $B_1$ is the amplified magnetic field in the precursor).

In the initial, linear stages of current-driven MFA the non-resonantly driven waves grow exponentially in time with characteristic 
fluctuation  scales much smaller than $r_{\rm g}$ of escaping particles
\citep{pelletier06}.
The maximum linear growth rate and the corresponding length scale are determined by
the return current, $j_{\rm CR}\propto u_{\rm s} p_{\rm max}^3 f(p_{\rm max})$, which
depends on the flux of escaping particles with $p\ga p_{\rm max}$ \citep{bell13}.
After the nonresonant mode becomes nonlinear (i.e., $\delta B/B_0 > 1$), 
for typical SNR shocks ($u_{\rm s}<0.1c$), the growth is dominated by the resonant mode and
the MHD turbulence continues to grow with fluctuations on increasingly larger scales 
up to $r_{\rm g}$, being limited by the advection time over which the shock sweeps the precursor plasma
\citep{marcowith10}.
Moreover, other processes such as the acoustic instability 
may operate simultaneously and lead to the growth of MHD turbulence on fluctuation scales larger 
than $r_{\rm g}$ \citep{schure12}. 
It has also been suggested that magnetic field fluctuations in shocks can grow on scales larger 
than $r_{\rm g}(p_{\rm max})$ 
in the presence of nonresonant circularly-polarized waves through the mean-field dynamo \citep{bykov11, roga12},
or through other microphysical and hydrodynamical instabilities within the shock precursor including the firehose, filamentation, and acoustic instabilities \citep[e.g.][]{reville12,beresnyak09,dru12,schure12,capri13}.

If MFA were confined only to a narrow region ($\ll l_{\rm max}$) close to the shock or occurred only downstream of the shock, 
the highest energy CRs with large diffusion lengths could not be accelerated efficiently,
and $p_{\rm max}$ would  still be limited by the background magnetic field, 
$B_0$, instead of the amplified field, $B_1$.
In most recent discussions CRs drive shock precursor MFA at rates that depend on the electric current associated with particle escape ahead of the shock \citep[e.g.,][]{bell04}\footnote{But, see e.g., \citet{beresnyak09, dru12} for alternate views.}, although it remains to be understood if the CR current owing to the escaping {\it highest energy} CRs
far upstream is a able to generate magnetic turbulence spanning lengths comparable to their own diffusion lengths. We do not attempt to address that issue here, but, instead apply several previously proposed phenomenological models for MFA motivated by this idea, in order to compare their impact on DSA and associated nonthermal
emissions in evolving SNR shocks. 
Our particular aim in this regard is to evaluate the importance of the distribution of the amplified magnetic field
within the CR precursor, since various interpretations of MFA in the literature lead to different distributions.

One of the signature consequences of standard nonlinear DSA theory in strong shocks 
is the hardening of the CR spectrum compared to
test particle DSA theory at momenta approaching $p_{\rm max}$ from
below, along with a steepening of the spectrum at low momenta
\citep[e.g.,][]{capri10b}. 
That is, the predicted CR spectrum becomes concave between the injection
momentum and the upper, cutoff momentum.
This expected behavior does not, however, seem to be reflected
in observed nonthermal emissions, as discussed below. In fact, $\gamma$-ray emissions seen in some SNRs seem best explained if the high energy 
CR spectra are actually steeper than predicted with test particle DSA theory.  Influences of MFA in the nonlinear DSA theory have
been suggested as one remedy for this conflict. We will explore that issue in this work.

In fact a reduction of the CR acceleration efficiency and a steepening of the CR spectrum are potentially important consequences of MFA. 
These would result from increased rates of
so-called Alfv\'enic drift \citep[e.g.][]{vladimirov08,capri12, kang12}, 
if the mean magnetic field is enhanced along the shock normal on scales large compared to particle gyroradii. 
Resonantly excited Alfv\'en waves tend to drift along the mean field in the direction of CR streaming with respect to the background flow, so opposite to the CR number (pressure) gradient. 
Then the mean convective velocity of the scattering centers
becomes $\bold{u} + \bold{u_{\rm w}}$, 
where $\bold{u_{\rm w}}$ is the mean drift velocity of the scattering centers \citep{skill75, bell78, pz05}.
For spherically expanding SNR shocks, the CR pressure peaks at the shock location. 
Thus, resonant waves moving outward into the background medium dominate in the
upstream region, while inward propagating waves may dominate 
behind the shock.
Then in mostly quasi-parallel spherical shocks it would be expected that the radial wave drift speed 
is $u_{\rm w,1} \approx +v_{\rm A}$ in the upstream 
shock precursor
(where $v_{\rm A}=B/\sqrt{4\pi \rho}$ is the local Alfv\'en speed),
while $u_{\rm w,2} \approx - v_{\rm A}$ behind the shock \citep{skill75, zp12}.
It is conceivable, however, in the downstream region of the forward SNR shock
that the forward and backward moving waves could be nearly balanced there 
(i.e. $u_{\rm w,2} \approx 0$) as a result of shock-related instabilities \citep[e.g.,][]{jon93}.
On the other hand, we do not have a fully self-consistent model for the wave generation and amplification
via wave-particle and wave-wave interactions around the shock. As we will demonstrate below, significant post-shock
drift in an amplified field could strongly influence the resulting shock and CR properties \citep[e.g.,][]{zp12}. 
 To illustrate that point simply, we will consider models in which
either $u_{\rm w,2} \approx 0$ or $u_{\rm w,2} \approx - v_{\rm A}$ is adopted.

These drifting effects probably are not relevant in the absence of MFA, since for typical Type Ia SNRs propagating into the interstellar medium, 
the effects of Alfv\'enic drift can be ignored. The Alfv\'enic Mach number is large for fast shocks in the interstellar medium,
e.g., $M_{\rm A}=u_{\rm s}/v_{\rm A} \sim 200$ for $B_0\approx 5 \muG$.
However, if the magnetic field strength is increased by a factor of 10 or more 
in the precursor, Alfv\'enic drift may affect significantly DSA at such SNRs.

In the presence of fast Alfv\'enic drift due to efficient MFA,
the velocity jumps that the scattering centers experience across the shock
would become significantly smaller than those of the underlying flow \citep[e.g.,][] {bell78, schlick89}.
Since CR particles are isotropized in the mean frame of scattering centers rather than the underlying fluid,
the resulting CR spectrum becomes softer  than that predicted with the velocity
jump for the background flow (see, e.g., Equations (\ref{qs}) and (\ref{qt})  below). Then DSA extracts less energy from the shock flow,  because the rate
at which particles gain energy is reduced compared to the rate of particle escape downstream. Consequently,  there are fewer
of the most energetic CRs \citep[e.g.,][]{kang12}. For this reason Alfv\'enic drift has been pointed out as a means to obtain a CR energy spectrum steeper
than the conventional test-particle power-law for strong shocks, e.g. $N(E)\propto E^{-2.3}$ \citep[e.g.][]{morlino12}, 
as required to explain the observed $\gamma$-ray spectra  as a consequence of secondary pion decay in the GeV-TeV band of 
some young SNRs \citep{abdo10,acero10,acciari11,capri11, giordano12}.

Exploring this effect, \citet{capri12} recently presented a nonlinear DSA model to produce a spectrum of SNR accelerated CRs  that is steeper than
the test-particle power-law at strong shocks by specifically accounting for Alfv\'enic drift in strongly amplified magnetic fields.
He adopted a magnetic field amplification model in which the turbulent magnetic fields 
induced by CR streaming instabilities increase rapidly to a saturation level that is spatially 
uniform within the shock precursor.  His model targeted spherical SNR shocks, but was based on sequential, semi-analytic, steady state DSA solutions; 
our work below, similarly motivated, applies spherical, explicitly time evolving numerical models to the problem. 

Beyond Alfv\'enic drift, there is another potentially important property of CRs in SNR 
shocks that may lead to steeper spectra at the highest energies. 
First, it is expected that the highest energy particles may escape rapidly from the system
when the diffusion length becomes greater than the shock curvature radius, 
i.e., $l_{\rm max} \ga  r_{\rm s}(t)$.
In particular, note that the Hillas constraint given above corresponds, with Bohm diffusion, 
to the condition $l_{\rm max} \sim (1/3) r_{\rm s}$. 
This would steepen the CR spectrum with respect to the plane shock solution.
which, in turn, would reduce the charge current driving instabilities, thus reducing
the efficiency of particle scattering at the highest energies.
However, considering that for typical SNRs, $r_{\rm s}\sim 3-10$ pc, while 
$l_{\rm max}\sim 0.5{\rm pc}(E_{\rm max}/1{\rm PeV})$ 
in the case of efficient MFA in the upstream region,
more stringent conditions due to reduction of MHD turbulence should be imposed here.
As we described above in the discussion of MFA in the shock precursor,
it remains uncertain up to what upstream location the highest energy particles can generate 
turbulent wave fields that are strong enough to confine themselves around the shock 
via resonant scattering \citep{capri10a,dru11}.
Moreover, during the late Sedov-Taylor stage of SNRs evolving in partially ionized 
media wave dissipation due to ion-neutral collisions may weaken stochastic 
scattering on the relevant scales, facilitating free streaming of high energy CRs 
away from the SNR and out of the DSA process \citep{pz05, malkov11}.
In order to account for such effects,
we consider a free escape boundary located at $r_{\rm FEB}= (1.1-1.5) r_{\rm s}(t)$.

We mention for completeness that alternate and more complex approaches to explaining
the steep $\gamma$-ray spectra in SNRs have been suggested. For example, \citet{bkv13} have proposed recently that
the observed, steep $\gamma$-ray spectrum of Tycho's SNR could be explained by pion production from
the combined populations of CR protons accelerated by shocks propagating into an ISM including 
two different phases.

As noted earlier, multi-band observations of nonthermal emissions from radio to $\gamma$-ray provide a powerful tool 
to test theoretical modeling of nonlinear DSA at SNRs \citep[e.g.][]{bkv09,bkv12,capri11,kang11,morlino12}. 
For instance the radio spectrum, $F_{\nu} \propto \nu^{-\alpha}$, represents 
the energy spectrum of electrons, $N_{\rm e}(E)\propto E^{-r}$  
(with $r= 2\alpha+1$ and $E= \gamma_{\rm e} m_{\rm e} c^2)$. 
In a magnetic field of typical strength, $B\sim 100\mu$G, these electrons have 
a characteristic Lorentz factor, $\gamma_{\rm e} \sim \sqrt{\nu m_{\rm e} c/(e B)} \sim 10^3$,
for radio synchrotron emissions in the GHz band.
If the peak CR electron energy is determined by a balance between DSA and synchrotron energy losses, and we assume for simplicity a
steady shock, the X-ray synchrotron cutoff frequency
is determined primarily by the shock speed; namely,
$h\nu_c\approx 4.1{\rm keV} (u_{\rm s}/3000\kms)^2$.
However, under similar conditions for spherical, decelerating shocks, radiative cooling of the CR electrons within the SNR interior
leads to a volume-integrated electron spectrum steepened above a break energy that depends on the 
evolution of $u_{\rm s}(t)$ and $B(r,t)$.
Then the spatially unresolved, synchrotron radiation spectrum has a break
at $h \nu_{\rm br} \sim 0.12 {\rm keV} (t/300 {\rm yr})^{-2} (B_2/100\muG)^{-3}$
above which the photon spectral index, $\alpha$, increases by 0.5
 compared to the value without radiative cooling \citep{kej12}. 

The interpretation of the $\gamma$-ray spectrum is more complicated, since       
$\gamma$-ray emission can originate from both CR protons and CR electrons; namely, by way of the decay of 
neutral pions produced in $p-p$ interactions between CRs and the background
medium, and from inverse Compton (iC) scattering of the background radiation by CRs electrons
plus nonthermal electronic bremsstrahlung. 
The relative importance of the different components is governed by
several factors, including the magnetic field strength, the background density, 
the background radiation field, and the CR electron to proton ratio, $K_{\rm e/p}$.  Given these ingredients, it is clearly crucial to incorporate
MFA, Alfv\'enic drift and particle escape in predicting nonthermal radiation spectrum of SNRs. 

In \cite{kang13} (Paper I) phenomenological models for MFA, Alfv\'enic drift 
and particle escape were implemented in time-dependent nonlinear DSA simulations of CR protons and electrons 
at the forward shock of Sedov-Taylor SNRs. Electronic synchrotron and iC losses were also included in the evolution of the electron spectra. 
Paper I demonstrated the following points for the MFA model employed there:
1) If scattering centers drift along the shock normal at the Alfv\'en speed in highly amplified magnetic fields, 
the CR energy spectrum is steepened in evolving strong SNR shocks and the acceleration efficiency is significantly 
reduced.
2) Even with fast Afv\'enic drift, however, DSA can still be efficient enough to develop 
a substantial shock precursor and convert about 20-30\% of the SN explosion energy
into CRs. 
3) A CR proton spectrum steeper than $E^{-2}$ was obtained only when Alfv\'enic drift away from the shock
was included in both upstream and downstream regions of the shock.
4) The maximum energy of CR ions accelerated by SNRs can increase significantly over values predicted without MFA
only when the magnetic fields are amplified in a volume spanning the full diffusion length of the
highest energy particles. This length scale is larger than the gyroradius of those particles 
by a factor of $(c/3u_{\rm s})$ when Bohm diffusion is applied.
5) Since the high energy end of the CR proton spectrum is composed of the particles that are
injected in the early stages of shock evolution, the $\gamma$-ray emission spectrum near the high energy
cutoff depends on details of the time-dependent evolution of
the CR injection, MFA, and particle escape as well as the dynamical evolution of the SNR shock.
Steady shock solutions cannot capture these features properly.
The present paper revisits these issues through a wider range
of MFA models.

Because of such interdependencies between MFA and DSA,
a self-consistent picture of the full problem requires at the least time dependent
MHD simulations combined with a kinetic treatment of nonlinear DSA.
That work remains to be done.
As a step in this direction, we implemented in Paper I a prescription for MFA 
and the resulting magnetic field profile based on a simple treatment of \citet{capri12},
of resonant amplification of Alfv\'en waves by streaming CRs 
(see Equation (\ref{Bpre}) below).
In the present work we will include three additional recipes for the magnetic field profile in the
shock precursor, applying models of \citet{zp08} and \citet{marcowith10}.
Moreover, we consider here slightly different model parameters from Paper I.
We also present the nonthermal radiation spectra calculated using the simulated CR proton and electron
spectra along with the magnetic field strength and the gas density profiles.
Our main aim is to explore how wave-particle interactions affect the energy
spectra of CR protons and electrons in nonlinear DSA at SNRs, 
and their nonthermal radiation spectrum.

In the next section we describe the numerical method for the simulations we report, 
phenomenological models for some key plasma interactions,
and model parameters for the Sedov-Taylor blast wave initial conditions.  
Our results will be discussed in Section 3, 
followed by a brief summary in Section 4.

\section{ NEW DSA SIMULATIONS}

In this section we briefly describe the numerical code and the phenomenological
models for wave-particle interactions in DSA theory that we applied. Full
details of similar DSA simulations can be found in Paper I.

\subsection{CRASH Code for DSA}

We consider parallel shocks, in which the magnetic fields can be roughly decoupled
from the dynamical evolution of the underlying flow.
The pitch-angle-averaged phase space distribution function, $f(p)$, for CR protons 
and electrons can be described by the following diffusion-convection equation 
\citep{skill75}:
\begin{eqnarray}
{\partial g\over \partial t}  + (u+u_w) {\partial g \over \partial r}
= {1\over{3r^2}} {\partial \over \partial r} \left[r^2 (u+u_w)\right]
\left( {\partial g\over \partial y} -4g \right) \nonumber\\
+ {1 \over r^2}{\partial \over \partial r} \left[r^2 \kappa(r,y)
{\partial g \over \partial r}\right]
+ p {\partial \over {\partial y}} \left( {b\over p^2} g \right) ,
\label{dc}
\end{eqnarray}
where $g=fp^4$,
$y=\ln(p/m_{\rm p} c)$ is the logarithmic momentum variable,
and $\kappa(r,p)$ is the spatial diffusion 
coefficient\footnote{For later discussion it is useful to note in the
relativistic regime that the momentum and energy distributions are simply
related as $n(E) = 4\pi p^2 f(p)$ with $E = pc$.}. 
In the last term $b(p)=-dp/dt$ is the electronic combined synchrotron and 
iC cooling rate. For protons $b(p)=0$.

The basic gasdynamic conservation laws with additional terms for the CR pressure,
$P_{\rm CR}$, CR induced non-adiabatic heating,
and an isotropic magnetic pressure, $P_{\rm B}$, are solved using the spherical version of
CRASH (Cosmic-Ray Amr SHock) code \citep{kj06}.
The CR pressure is calculated self-consistently from the CR proton distribution function,
$g_{\rm p}(p)$, determined from the finite difference solution to equation (\ref{dc}).
The magnetic pressure is calculated according to our phenomenological 
models for MFA (see the following Section 2.2) rather than from direct solutions of the induction equation or MHD wave transport equations.

\subsection{Magnetic Field Amplification (MFA)}

CRs streaming across the gas subshock into the shock precursor are known to generate resonant and nonresonant waves
via streaming instabilities \citep[e.g.,][]{lucek00,bell04,zp08}.
In that event MHD perturbations of short wavelengths ($\lambda \ll r_{\rm g}$) 
being advected into the shock precursor are amplified by
the return charge current induced by escaping high energy CR particles, 
$j_{\rm CR} \sim e \pi u_{\rm s}  p^3_{\rm max} f(p_{\rm max})$ 
\citep{bell13}.
The linear nonresonant instability grows fastest on scales, 
$k^{-1}_{\rm max}\sim Bc/ (2\pi j_{\rm CR}$) at
a rate, $\Gamma_{\rm max}\sim (j_{\rm CR}/c) \sqrt{\pi/\rho}$.
So, the amplification of MHD turbulence via nonresonant interactions depends on the flux
of escaping particles, which, in turn, is governed by the efficiency of the CR acceleration and the shape of the CR spectrum at large momenta. 
As the instability enters the nonlinear regime ($\delta B\sim B_0$) during
passage through the precursor,
saturation processes start to limit the growth, and previously sub-dominant resonant interactions become dominant.
This leads to a linear growth of magnetic fluctuations and extension to larger
scales \citep{pelletier06}.
As noted in the introduction, there may also exist other types of instabilities and dynamo mechanisms
leading to the growth of MHD turbulence on scales larger 
than $r_{\rm g}$ \citep[e.g.,][]{schure12}.
A full understanding of complex interplay between MFA and DSA would require
MHD simulations combined with nonlinear DSA
in which the return current is calculated self-consistently from the accelerated CR spectrum
\citep[see][for a test-particle treatment]{marcowith10}.
Our more limited objective in the present study is to explore broadly the impact
of the resulting MFA profile.
Consequently, we implement four heuristic models for MFA in the precursor designated {\bf{M1}} - {\bf{M4}},
each established by simple applications of MFA
and compare their consequences in DSA within model SNR shocks.

MFA model {\bf{M1}}: \citet{capri12} has shown for strong shocks with $M_{\rm s}\gg 1$ and $M_{\rm A}\gg 1$,
that the strength of the turbulent magnetic field amplified via resonant Alfv\'en waves excited by CR
streaming instabilities can be approximated in terms of the flow speed (compression) 
within the shock precursor, as
\begin{equation}
{B(r)^2 \over B_0^2} = 1 + (1-\omega_{\rm H})\cdot {4\over 25}M_{\rm A,0}^2 { {(1-U(r)^{5/4})^2}\over U(r)^{3/2}},
\label{Bpre}
\end{equation}
where $U(r)= [u_{\rm s}-|u(r)|]/u_{\rm s} = \rho_0/\rho(r)$ is the normalized flow speed with respect to
the shock, and
$M_{\rm A,0}= u_{\rm s}(t)/v_{\rm A,0}$ is the Alfv\'enic Mach number for the instantaneous
shock speed with respect to the far upstream Alfven speed, 
$v_{\rm A,0}= B_0/ \sqrt{4\pi \rho_0}$.
We hereafter designate this magnetic field profile as model {\bf{M1}}
and use the subscripts `0', `1', and `2' to denote
conditions far upstream of the shock, immediately upstream and 
downstream of the subshock, respectively.
The factor $(1-\omega_{\rm H})$ accounts for local wave dissipation and
the ensuing reduction of MFA; i.e., a fraction, $\omega_{\rm H}$, of the
energy transferred from CR streaming to MHD waves is dissipated
as heat in the plasma by way of nonlinear damping processes. Some damping is
likely; we arbitrarily set $\omega_{\rm H} = 0.5$ as a reasonable estimate.
In Paper I, this {\bf{M1}} recipe was adopted to represent qualitatively the MFA process 
in the shock precursor.

As we will show below, Alfv\'enic drift of scattering centers
upstream at the
local Alfv\'en speed, $v_{\rm A} = B(r)/\sqrt{4\pi\rho(r)}$,
along the shock normal can steepen the CR spectrum significantly in the presence of MFA.
On the other hand, the magnetic field in the {\bf{M1}} MFA model (Equation (\ref{Bpre})) 
increases gradually through the shock precursor
from $B_0$ at the FEB to $B_1$ at the subshock (see also Figure 1, below).
Consequently, the drift speed increases slowly through the precursor, so that
the highest energy CRs, which diffuse on scales
$\kappa(p_{\rm max})/u_{\rm s} \sim L$,
are scattered mostly by waves with the relatively slow drift speed, $v_{\rm A,0}$.  
(The length $L$, defined below, measures the full width of the precursor.)
That makes Alfv\'enic drift and associated CR spectral steepening ineffective
at the high energy end of the CR spectrum.

MFA model {\bf{M2}}: On the other hand, \citet{capri12} pointed out that Equation (\ref{Bpre}) does not
account for several important effects, such as excitation of the nonresonant
streaming instability that can rapidly amplify the field at the leading edge of the precursor. 
To allow for such influences, he proposed an alternative simple MFA profile in which
the entire upstream, precursor
region, $0 < (r-r_{\rm s}) < L$, has the saturated
magnetic field, $B_1$; i.e.,
\begin{equation}
B(r) = B_1,
\label{Bpre2}
\end{equation}
where $B_1$ is calculated according to Equation (\ref{Bpre}).
We hereafter designate that MFA profile as model {\bf{M2}}.
With the help of this saturated, uniform precursor magnetic field profile, 
\citet{capri12} obtained a CR energy spectrum steeper than $E^{-2}$ from
nonlinear DSA calculations of SNRs.

MFA model {\bf{M3}}: \citet{zp08} carried out MHD simulations following the evolution of the
nonresonant current instability through shock precursors. Their MFA profile
was approximately exponential (see their Figure 3). We adapt this behavior into the simple form
(designated, hereafter, model {\bf{M3}})
for $0 \le(r-r_{\rm s})\le L$,
\begin{equation}
B(r) = B_0 +  B_1 \cdot ({B_1 \over {\delta B_0}})^{-(r - r_{\rm s})/L},
\label{Bpre3}
\end{equation}
where $\delta B_0=0.01$ is an assumed, arbitrary strength of the initial background
magnetic field perturbations and $L$ is the distance from the shock to the upstream boundary
(see Section 2.4).
Again, for the maximum magnetic field strength immediately upstream of
the shock, we adopted $B_1$ calculated according to Equation (\ref{Bpre}).

MFA model {\bf{M4}}: As a fourth model for MFA in the precursor, we adopt a linear magnetic strength profile as in \citet{marcowith10} (see their Figure 2):
so for $0\le (r-r_{\rm s})\le L^*$,
\begin{equation}
B(r) = B_1 - (B_1-B_0)\cdot {{r-r_{\rm s}}\over L^*}
\label{Bpre4}
\end{equation}
where $L^*=0.9L$ is used somewhat arbitrarily. For $(r-r_{\rm s})>L^*$, $B(r)=B_0$.  
This model represents an exponential growth to $\delta B \sim B_0$ on a short time scale
at $r\sim L^*$ by nonresonant modes followed by a linear growth to $B_1$ through combined 
nonresonant and resonant mode amplification.
The profiles {\bf{M1}} through {\bf{M4}} can be compared in Figure 1.

In the case of ``strongly modified'' shocks (e.g., $U(r_{\rm s}) \ll 1$), magnetic field
energy density may increase to a significant fraction of the
upstream kinetic energy density, $(1/2)\rho_0 u_{\rm s}^2$, which is not compatible with observations.  \citet{vbk05} have found postshock
magnetic pressures, $B_2^2/(8\pi)$, 
in several young, shell-type SNRs that are all several \% of the upstream ram pressures, $\rho_0 u_{\rm s}^2$ .
Using this for guidance in the simulations we restricted the amplification factor within the precursor by the condition that
\begin{equation}
(B_1^2/8\pi)\la (B_{\rm sat}^2/8\pi)\equiv 0.005\rho_0 u_{\rm s}^2.
\label{Bsat}
\end{equation}
For typical Type Ia SNRs in the warm ISM with $n_H=0.3 {\rm cm}^{-3}$ (see the section 2.5 and Table 1),
the unmodified Sedov-Taylor solution is $U_{ST}=5.65\times 10^3 \kms(t/t_o)^{-3/5}$,
so $B_{\rm sat} = 168\muG (t/t_o)^{-3/5}$. 
In the case of a moderate precursor with $U_1\approx 0.8$ or $\rho_1/\rho_0\approx 1.25$, 
for $\omega_H= 0.5$ and $B_0=5 \muG$ Equation (\ref{Bpre}) gives
$B_1 \approx 0.081 M_{A,0} B_0\approx 136 \muG (t/t_o)^{-3/5}$.
So unless the shock is modified quite strongly, that is, $U_1<0.8$, $B_1$ should not exceed
the saturation limit $B_{\rm sat}$ for the warm ISM models considered here.
We note that the relation (\ref{Bsat}) was found observationally for young SNRs,
so its validity for much slower shocks at late Sedov stage has not been established.

Note that Equation (\ref{Bsat}) is equivalent to 
\begin{equation}
M_{\rm A,1}  = u_1/v_{\rm A,1} \ga \frac{10} {\sqrt{\sigma_1}},
\label{masat}
\end{equation}
where $\sigma_1 = u_0/u_1$ measures compression through
the precursor. 
This is useful in evaluating the influence of Alfv\'enic drift within the 
precursor (see Equations \ref{qs})-(\ref{qtp})).
 
We assume that the turbulent magnetic field is isotropic as it comes into the subshock
and that the two transverse components are simply compressed across the subshock. 
The immediate postshock field strength is estimated by
$B_2/B_1=\sqrt{1/3+2/3(\rho_2/\rho_1)^2}$.
If subshock compression is large, $B_2/B_1 \approx 0.8( \rho_2/\rho_1)$.
From Equation (\ref{Bsat}) this leads, as a rule of thumb, roughly to $B_2^2/(8\pi)/(\rho_0 u_{\rm s}^2) \la 3$\% (see Figure 6).
It is not well understood how the magnetic fields diminish downstream in the flow behind the forward shock
\citep[e.g.][]{pohl05}.
We assume for simplicity that the postshock
field strength behaves as $B(r) = B_2 \cdot \left[\rho(r)/ \rho_2 \right]$
for $r < r_{\rm s}$.  

\subsection{Alfv\'enic Drift}

As noted earlier, the Alf\'ven waves generated by the streaming instability drift along the local mean magnetic field with respect to the 
background plasma flow in the direction opposite to the CR gradient. Ahead of the shock these waves would propagate into the background
medium; behind a spherical shock those waves would propagate towards the SNR interior. Since scattering isotropizes the CRs 
with respect to the scattering centers, the effective velocity difference that the particles
experience across the shock is reduced, if those waves dominate CR scattering.  The reduced velocity jump softens (steepens) the CR momentum
spectrum compared to the test-particle result for a
shock without a precursor or Alfv\'enic drift, $q_0 = 3u_0/(u_0 - u_2)$, where 
$q=-\partial \ln f/\partial \ln p$. 

In a CR modified shock with a precursor and Alfv\'enic drift the slope of the momentum distribution function
is momentum dependent, reflecting the variation with momentum of the particle
diffusion length and the different flow conditions sampled by the particles
as a result. Near the momentum extremes the slopes can be estimated for
steady, plane shocks as:
\begin{equation}
q_{\rm s} \approx {{3(u_1+u_{\rm w,1})}\over {(u_1+u_{\rm w,1})-(u_2+u_{\rm w,2})}},
\label{qs}
\end{equation}
for the low energy particles just above the injection momentum ($p\sim p_{\rm inj}$), and
\begin{equation}
q_{\rm t} \approx {{3(u_0+u_{\rm w,0})}\over {(u_0+u_{\rm w,0})-(u_2+u_{\rm w,2})}},
\label{qt}
\end{equation}
for the highest energy particles just below the cutoff; i.e., $p \la p_{\rm max}$.
Here $u_0= - u_{\rm s}$, $u_1=-u_{\rm s}/\sigma_1$, $u_2= -u_{\rm s}/\sigma_2$. 
To make
the point of the impact of Alfv\'enic drift with minimal complication, we assume for drift velocities simply
$u_{\rm w,0}=+v_{\rm A,0}$, $u_{\rm w,1}=+v_{\rm A,1}$, and $u_{\rm w,2}=-v_{\rm A,2}$,
with $\sigma_1=\rho_1/\rho_0$, $\sigma_2=\rho_2/\rho_0$.
The local Aflv\'en speed is defined as $v_{\rm A,*}=B_*/\sqrt{4\pi \rho_*}$ (where $*=0$, 1, 2).
The steepening of CR spectrum due to Alfv\'enic drift can obviously be ignored for high Alfv\'enic Mach numbers, 
$ M_{\rm A,*}= u_{\rm s}/v_{\rm A,*} \gg1 $.
As noted earlier, postshock Alfv\'enic drift is frequently assumed to
vanish; i.e., $u_{\rm w,2} = 0$, based on the argument that
postshock turbulence is likely to be balanced \citep[e.g.,][]{jon93}.
Here we want to emphasize, on the other hand that the shocks in this discussion
are not really steady, plane shocks, but evolving, spherical shocks.
There should generally be a strong CR gradient behind a spherical
shock to drive a streaming instability, so that one could reasonably expect $u_{\rm w,2} < 0$
\citep[e.g.,][]{zp12}. 
If this
happens, it could significantly influence the CR spectrum.

To see the dependencies on shock modification and Alfv\'enic drift
more clearly, suppose $|u_{\rm w,*}/u_*|\ll 1$ (where $* = 0,1,2$) and $(\sigma_1 - 1)/(\sigma_2 - 1) \ll1$, so that we can
expand the corrections of $q_{\rm s}$ and $q_{\rm t}$ compared to the fiducial slope, $q_0 = 3\sigma_2/(\sigma_2 - 1)$.
Then keeping only lowest order corrections to $q_0$ we can write,
\begin{equation}
q_{\rm s} \approx q_0 \left[ 1 + \frac{\sigma_1 - 1}{\sigma_2 - 1} + \frac{\sigma_1}{\sigma_2 - 1}\left(\frac{u_{\rm w,2}}{u_2} - \frac{u_{\rm w,1}}{u_1}\right)\right],
\label{qsp}
\end{equation}
and
\begin{equation}
q_{\rm t} \approx q_0 \left[ 1 + \frac{1}{\sigma_2 - 1}\left(\frac{u_{\rm w,2}}{u_2} - \frac{u_{\rm w,0}}{u_0}\right)\right].
\label{qtp}
\end{equation}
Compression through the precursor steepens $q_{\rm s}$ compared to $q_{\rm t}$, illustrating
the concavity of CR spectra usually predicted by nonlinear DSA. 
The slope, $q_0$, set by the full compression, $\sigma_2$,
is, of course, flatter than the slope in an unmodified shock of similar sonic Mach number.
One can also see that the slopes, $q_{\rm s}$ and $q_{\rm t}$, are increased compared to the case with no Alfv\'enic drift,
if scattering centers drift upwind ahead of the shock and downwind
behind the shock. 
Suppose for the moment that the wave drift speed scales simply with the local Alfv\'en speed,
that $B_1 \gg B_0$ (see Equation (\ref{Bpre}) and Figure 6) and for
simplicity that $B_2/B_1 \sim \rho_2/\rho_1$. 
Then the influence of drift just upstream of the subshock is large compared to that just inside the FEB,
since $|(u_{\rm w,1}/u_1)/(u_{\rm w,0}/u_0)| \sim (B_1/B_0)\sigma_1^{1/2}\gg 1$.
One can also see that the downstream drift has greatest influence on both $q_{\rm s}$ and $q_{\rm t}$, 
because $(u_{\rm w,2}/u_2)/|(u_{\rm w,1}/u_1|) \sim (\sigma_2/\sigma_1)^{3/2}\gg 1$ and 
$(\sigma_1 - 1)/(\sigma_2 - 1) \ll1$ in the simulations we present here.
We note for clarity that the MFA constraint given in Equation (\ref{masat}) limits 
the preshock Alfv\'enic drift correction term, $|(u_{\rm w,1}/u_1)|$, in these simulations 
to $|(u_{\rm w,1}/u_1)| \la 0.1 \sqrt{\sigma_1}$.

From this discussion it should be obvious that the presence and nature of
downstream Alfv\'enic drift has a very important effect on the DSA outcomes.
It is also clear under these circumstances
that these expressions always satisfy $q_{\rm s} \ge q_{\rm t}$; i.e., measured
at the extremes the nonlinear spectra will remain concave, at least in a
steady, plane shock, even when Alfv\'enic drift
is included on both sides of the shock. On the other hand, spherical
shocks are not steady, and relative postshock flow speeds increase 
with distance downstream, opening up a wider range of possible outcomes.
We shall see, in fact, that in our SNR simulations the CR spectra near
the maximum, cutoff momentum can be significantly flatter than one would
predict from the above relations. That feature does result from strong
evolution of the shock properties at early times that influences subsequent
evolution. Shock and CR properties in spherical blast waves are not a simple superposition of
intermediate properties at the final time.

When Alfv\'enic drift has been included in DSA models, it has been customary
to assumed the drift speed is the local Alfv\'en speed along
the shock normal, $v_{\rm A} = B/\sqrt{4\pi\rho}$, based on the total magnetic
field strength.
However, when the fields become strongly amplified by streaming
instabilities the mean field direction is less clear, even when the
upstream field is along the shock normal \citep[e.g.,][] {reville13}.
Then the drift speed should be reduced compared to the Alfv\'en speed
expressed in terms of the total field strength.
In order to allow for this we model the local effective Alfv\'enic drift speed simply as
\begin{equation}
 v_{\rm A}(r) = { {B_0 + f_{\rm A} [B(r)-B_0]} \over \sqrt{4\pi \rho(r)}},
\label{vA}
\end{equation}
where the parameter $f_{\rm A} \le 1$ is a free parameter \citep{pzs10,lee12}. 
For the simulations presented here $f_{\rm A} = 0.5$ whereever Alfv\'enic drift is active.
The default in our models turns off Alfv\'enic drift in the post shock flow;
i.e., we set $u_{\rm w,2} = 0$.
On the other hand, to allow for possible influences of postshock
streaming in these spherical shocks we also consider the case of 
$u_{\rm w,2}(r) = -v_{\rm A}(r)$.
Those models are identified with the subscript tag, '${ad}$' (see Table 1).
All the simulations presented here apply Alfv\'enic drift upstream of the shock, with  $u_{\rm w,1}(r)=+v_{\rm A}(r)$.

\subsection{Recipes for Particle Injection, Diffusion, and Escape}

We apply a thermal leakage model for CR injection in which
only suprathermal particles in the tail of the thermal, Maxwellian distribution 
above a critical rigidity are allowed to cross the shock from downstream to upstream.
CR protons are effectively injected above a prescribed injection momentum,
$p_{\rm inj} \approx 1.17 m_{\rm p} (u_{\rm s}/\sigma_2) (1+ 1.07 \epsilon_B^{-1})$,
where $\epsilon_B$ is an injection parameter defined in \citet{kjg02}.
We adopt $\epsilon_B=0.2-0.215$ here, which leads to the
injected proton fraction, $\xi = n_{\rm cr,2}/n_2 \la 10^{-4}$. 
In Paper 1, $\epsilon_B=0.23$ was adopted, which led a higher injection fraction 
than we allow here and higher
CR acceleration efficiencies, as well, that were nearly in the saturation regime.
Electrons are expected to be injected
with a much smaller injection rate than protons, since suprathermal electrons have much smaller
rigidities at a
given energy. Some preacceleration process is likely to control this  \citep{reynolds08}.
Since this physics is still poorly understood, we follow the common
practice of fixing the injected CR electron-to-proton ratio to a small number,
$K_{\rm e/p} \sim 10^{-4}-10^{-2}$ \citep[e.g.,][]{morlino12}.
Because of the small number of particles, the electronic CR component is dynamically
unimportant; we neglect its feedback in these simulations.

In these simulations injected proton and electron CRs 
are accelerated in the same manner at the same rigidity, $R=pc/Ze$.
For the spatial diffusion coefficient, we adopt a Bohm-like momentum dependence ($\kappa_B \sim (1/3) r_{\rm g} v$) 
with flattened non-relativistic dependence to reduce computational
costs \citep{kj06}. Since acceleration to relativistic energies is generally
very quick, this low energy form has little impact our results. In particular
we set
\begin{equation}
\kappa(r,p) = \kappa_{\rm n} ({B_0 \over B_{\parallel}(r)}) \cdot ({p \over m_{\rm p}c})
\cdot K(r), 
\label{Bohm}
\end{equation}
where $\kappa_n= m_{\rm p} c^3/(3eB_0)= (3.13\times 10^{22} {\rm cm^2s^{-1}}) B_0^{-1}$
($B_0$ is expressed in units of microgauss) and
the parallel component of the local magnetic field, $B_{\parallel}(r)$, is prescribed by 
our MFA models discussed above.
The function $K(r) \ge 1$ is intended to represent a gradual decrease in scattering
efficiency relative to Bohm diffusion (so, $\lambda_{\rm s} > r_{\rm g}$) 
upstream of the subshock due to such influences as predominantly sub-gyro-scale turbulent
fluctuations. All the simulations set $K(r) = 1$ for $r < r_{\rm s}$ (postshock region).
But, except for one model (WM1$_{\rm Bohm}$), where $K(r) = 1$, we use for $r\ge r_{\rm s}$,
\begin{equation}
K(r)=\exp[{{c_k\cdot(r-r_{\rm s})}\over r_{\rm s}}]. 
\end{equation}
The numerical factor, $c_k=20$, is chosen somewhat arbitrarily.
It can be adjusted to accommodate a wide range of effects \citep{zp12}.

In these simulations we explicitly provide for particle escape from the system
by implementing a so-called
``Free Escape Boundary'' or ``FEB'' a distance, $L$, upstream of the shock.
That is, we set $f(r_{\rm FEB},p)=0$ at $r_{\rm FEB}(t)=r_{\rm s} + L= (1+\zeta) r_{\rm s}(t)$, where $ \zeta=0.1-0.5$.
Once CRs are accelerated to high enough momenta, $p_{\rm max}$, that the diffusion length 
$l_{\rm max} = \kappa(p_{\rm max})/u_{\rm s} \sim L(t)$, 
the length $L$ becomes the effective width of the shock precursor. 
For $\zeta=0.1$ and the shock radius, $r_{\rm s}=3$ pc, the distance of the FEB from the shock is $L=0.3$ pc,
which is comparable to the diffusion length of PeV protons if $B\sim 50 \muG$ and $u_{\rm s}\sim 6700 \kms$.
Note that, for $c_k=20$ and $\zeta=0.1~(0.25)$, 
the value of $K(r)$ increases from unity at the shock to $e^2=7.4 ~(e^5=150)$ 
at the FEB.

The time-integrated spectrum of particles escaped from the shock from the start of
the simulation, $t_{\rm i}$ to time $t$ is given by
\begin{equation}
\Phi_{\rm esc}(p)=  \int_{t_{\rm i}}^t  4\pi r_{\rm FEB}(t)^2  [ \kappa\cdot |{{\partial f} \over {\partial r}}|]_{r_{\rm FEB}} ~{\rm dt }.
\label{Phi}
\end{equation}

Finally, we note that the rate of non-adiabatic gas heating due to wave dissipation in the
precursor is prescribed as 
$W(r,t)= -  \omega_{\rm H} \cdot v_{\rm A}(r) {\partial P_{\rm CR}/\partial r }$,
where a fiducial value of $\omega_{\rm H}=0.5$ was assumed in these simulations.

\subsection{Sedov-Taylor Blast Waves}

For a specific shock context we consider a Type Ia supernova explosion with the ejecta mass,
$M_{\rm ej}=1.4 M_{\sun}$, expanding into a uniform ISM.
All models have the explosion energy, $E_{\rm o}=10^{51}$ ergs.
Previous studies have shown that
the shock Mach number is one of the key parameter determining the evolution 
and the DSA efficiency \citep[e.g.,][]{kang10},
so two phases of the ISM are considered:
the {\it warm phase} with $n_{\rm H}=0.3{\rm cm}^{-3}$ and $T_0=3\times 10^4$K (`W' models),
and the {\it hot phase} with $n_{\rm H}=0.01~ {\rm cm}^{-3}$ and $T_0=10^6$K (`H' models).
The background gas is assumed to be completely ionized with the mean molecular weight,
$\mu=0.61$.
The background magnetic field strength is set to be $B_0=5\muG$. 
For the warm ISM ('W') models the upstream Alfv\'en speed is then $v_{\rm A,0}=16.8 \kms$. 
The associated shock Alfv\'en Mach number is
then $M_{\rm A,0}\approx 180 (u_{\rm s}/3000 \kms)$.
For the hot ISM ('H') models, $v_{\rm A,0}=183 \kms$, and $M_{\rm A,0}\approx 16.4 (u_{\rm s}/3000 \kms)$.

The model parameters for the DSA simulations are summarized in Table 1.
The second and third characters of the model name in column one indicate the MFA model 
profiles; namely,
'M1' for the velocity-dependent profile given by Equation (\ref{Bpre}),
'M2' for the uniform, saturated profile given by Equation (\ref{Bpre2}),
'M3' for the exponential profile given by Equation (\ref{Bpre3}),
and 'M4' for the linear profile given by Equation (\ref{Bpre4}).
The downstream default in all models sets $u_{\rm w,2}=0$; i.e., downstream Alfv\'enic drift is turned off.
Where downstream Alfv\'enic drift is operating; i.e., where $u_{\rm w,2}=-v_{\rm A,2}$, the
model labels include the subscript '${ad}$'.
In two models, WM1$_{\rm li}$ and HM1$_{\rm li}$ models, the injection rate is lowered by
reducing slightly the injection parameter from $\epsilon_B=0.215$ to
$\epsilon_B=0.2$.
The WM1$_{\rm Bohm}$ model is the same as the WM1 model except that the function $K(r)=1$ for $r>r_{\rm s}$,
instead of the defaut form given in Equation (\ref{Bohm}).
Models WM1$_{\rm feb2}$ and WM1$_{\rm feb3}$ are included to study the effects of different FEB locations; namely,
$L=r_{\rm FEB}-r_{\rm s}=0.25r_{\rm s}$ and $L = 0.5r_{\rm s}$, respectively.
For all models, $f_{\rm A}=0.5$ is adopted for the Alfv\'en drift parameter
and $\omega_{\rm H}=0.5$ for the wave dissipation parameter.

The physical quantities are normalized, both in the numerical code and in
the plots below, by the following constants:
$r_{\rm o}=\left({3M_{\rm ej}/4\pi\rho_{\rm o}}\right)^{1/3}$,
$t_{\rm o}=\left({\rho_{\rm o} r_{\rm o}^5 / E_{\rm o}}\right)^{1/2}$,
$u_{\rm o}=r_{\rm o}/t_{\rm o} $,
$\rho_{\rm o} = (2.34\times 10^{-24} {\rm g cm^{-3}}) n_{\rm H}$,
and $P_{\rm o}=\rho_{\rm o} u_{\rm o}^2$.
For $r = r_{\rm o}$ the mass swept up by the forward shock equals the ejected mass, $M_{\rm ej}$.
For the warm ISM models, $r_{\rm o}=3.18$pc and $t_{\rm o}=255$ years, while
for the hot ISM models, $r_{\rm o}=9.89$pc and $t_{\rm o}=792$ years.

The true Sedov-Taylor (ST hereafter) dynamical phase of SNR evolution
is established only after the reverse shock is 
reflected at the explosion center. So, the dynamical evolution of young SNRs is much more
complex than that of the ST similarity solution that we adopt for a simple initial
condition at $t_{\rm i}$.
In particular we start each simulation from the ST similarity solution, $r_{\rm ST}/r_{\rm o}=1.15(t/t_{\rm o})^{2/5}$,
without the contact discontinuity and the reverse shock.
Although the early dynamical evolution of the model SNRs is not accurate in these simulations,
the evolution of the forward shock is still qualitatively representative \citep[e.g.,][]{dohm96}.
Our main goal is to explore how
time dependent evolution of MFA, Afv\'enic drift and FEB affect
the high energy end of the CR spectra rather than to match the properties of a
specific SNR.
On the other hand, since the highest energy CRs generally correspond to those
injected into the DSA process at very early times, 
before the ST stage, we begin the calculations 
at $t_{\rm i}/t_{\rm o}=0.2$. We then follow the evolution of the forward
shock to $t/t_{\rm o}=10$, which corresponds roughly to the beginning of the true ST evolutionary phase.

The spherical grid used in the simulations expands in a comoving way with the forward shock \citep{kj06}.
Continuation conditions are enforced at the inner boundary, which is located at $r/r_{\rm s}=0.1$ at the start of each simulation and moves outward along with the forward shock.
At the outer boundary the gasdynamic variables are continuous, while
the CR distribution function is set by the FEB condition.

\subsection{Emissions}

The nonthermal radio to $\gamma$-ray emissions expected in these simulations
from CR electrons and from CR proton secondary products, along with
thermal Bremsstrahlung were computed using an updated
version of the COSMICP (renamed AURA) code described in \citet{ekjm11}. 
We include only information essential for clarity here. 
Specifically we include for CR electrons synchrotron, iC and Bremsstrahlung. 
The iC emissions properly account for electron recoil
in the Klein-Nishina limit. Hadronic interactions include inelastic proton-proton
and photon-proton collisions. The low energy proton-proton cross-section was updated using \citet{knttt06}. Helium and heavy ion contributions are
ignored. The photo-pion production rates and iC rates
are based on a background radiation field with a total energy density 1.04 eV cm$^{-3}$ 
including the cosmic microwave background, plus contributions from cold dust,
old yellow and young blue stars, as described in \citet{ekjm11}.

\section{DSA SIMULATION RESULTS}

\subsection{CR Properties}

\begin{figure}
\vspace{+1cm}
\begin{center}
\includegraphics[scale=0.85]{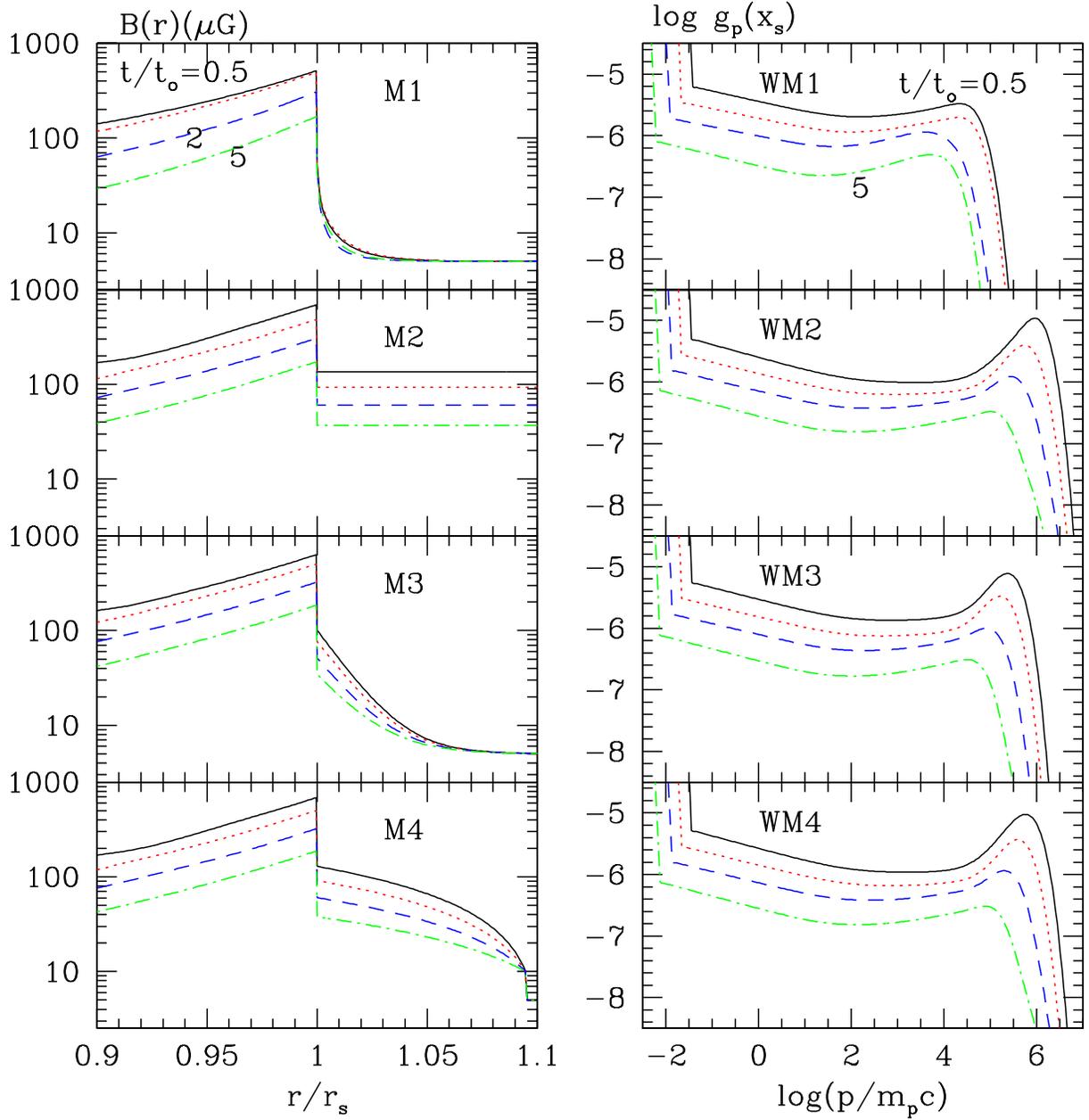}
\end{center}
\vspace{-0.0cm}
\caption{
Magnetic field profile, $B(r)$, and the CR proton spectrum at the subshock location, 
$g_{\rm p}(x_{\rm s})$, are shown 
at $t/t_{\rm o}=0.5$ (black solid lines), 1 (red dotted), 2 (blue dashed), and 5 (green dot-dashed)
in WM1 (Equation~[2]), WM2 (Equation~[3]), WM3 (Equation~[4]) and WM4 (Equation~[5]) models (from top to bottom). 
See Table 1 for model parameters.
The radial coordinate is normalized with the shock radius, $r_{\rm s}(t)$ at a given epoch.
Here $t_{\rm o}=255$ yr.
}
\label{fig1}
\end{figure}

Figure \ref{fig1} shows at times $t/t_0 = 0.5, 1, 2, 5$ the amplified magnetic field in the radial coordinate
normalized by the shock radius, $r/r_{\rm s}(t)$, and the distribution function of CR
protons at the subshock, $g_{\rm p}(x_{\rm s})$, for warm ISM simulations with the four different
MFA models: WM1, WM2, WM3, and WM4.
In the figures below the momentum is expressed in units of $m_pc$ and
the particle distribution function is defined in such a way 
that $\int_0^{\infty} 4\pi p^2 f_0(p)dp = n_H$ far upstream.
So both $f(p)$ and $g(p)=f(p)p^4$ are given in units of $n_H$,
while the volume-integrated distribution functions 
$F(p) = 4\pi \int f(r,p) r^2 dr$ and $G(p)=p^4 F(p)$ are given in units of $n_H r_o^3$.

During the early stage ($t/t_o<1$) the precursor grows to a time-asymptotic structure and
the magnetic fields are amplified rapidly to saturation.
Afterwards, the value for $B_1$ declines as $B_1 \propto M_{A,0}\propto (t/t_o)^{-3/5}$,
as the shock slows down in time.
The time evolution of MFA along with other measures,
including the precursor modification and the postshock CR pressure, will be discussed below in Figure 6.
The distance from the subshock to the FEB location is the same,
$L=0.1r_{\rm s}$, in these models, but the magnetic field profiles differ in shape.
The ``effective width'' of the magnetic field precursor (measured, say, by
the half-power width) would increase through the sequence {\bf{M1}, \bf{M3}, \bf{M4}, \bf{M2}}.
One can see that the DSA efficiency increases among the WM1-4 models 
in the same order at early times ($t/t_0 \la 1$).
This efficiency influence is reflected in Figure \ref{fig1} most clearly
in the variation in $p_{\rm p,max}$. 
This demonstrates that $p_{\rm p,max}$ is determined not only by the strength of $B_1$ but
also by the width of the magnetic field precursor.

As the acceleration proceeds from $t_{\rm i}/t_{\rm o}=0.2$, the maximum momentum increases and
reaches its highest value at $t/t_{\rm o}\sim 0.5$, depending in detail on the profile of $B(r,t)$.
Afterwards, $p_{\rm p,max}$ decreases in time as the shock slows down and MFA becomes
weaker (as shown in the figure).
For the {\bf M2}, {\bf M3} and {\bf M4} models, 
the highest proton energy reaches $E_{\rm p,max}\sim 1$ PeV 
at $t/t_{\rm o}\sim 0.5$.
For the {\bf M1} model, the magnetic field precursor is too narrow to provide a significant
enhancement of DSA over that from the upstream field, $B_0$, so the proton energy 
increases only to $E_{\rm p,max}\sim 0.05$ PeV.
 
Intuitively, the high momentum end of $g_{\rm p}(r_{\rm s})$ 
is obviously populated by particles injected 
during the earliest period of very efficient DSA, when the shock was especially
strong.
At the beginning of the simulations ($t_{\rm i}/t_{\rm o}=0.2)$, 
the shock is fast and strong with $u_s\approx 1.5\times 10^4 \kms$,
$M_{\rm s,0}\approx 580$, and $M_{\rm A,0}\approx 890$,
so the initial slope of $f_{\rm p}(r_{\rm s})$ starts with $q=4$.
As the precursor develops, the CR spectrum becomes concave ($dq/d\ln{p} < 0$).
Although Alfv\'enic drift increases both $q_{\rm s}$ and $q_{\rm t}$ as
MFA proceeds in the precursor, these effects are small initially because of
large values of $M_{\rm A,1}$.
In the WM2 model, with a uniform precursor magnetic field, for example,
$M_{\rm A,1}\approx M_{\rm A,0}\approx20$ for $B_1=200\muG$ and $u_s\approx 1.5\times 10^4 \kms$,
resulting in only small corrections to $q_{\rm s}$ and $q_{\rm t}$ (see Equations 
(\ref{qsp})-(\ref{qtp})).
Even in the later stage ($t_{\rm i}/t_{\rm o}\ga 1$), $M_{\rm A,1}>10$ according to
Equation (\ref{masat}), so steepening of CR spectra due to Alfv\'enic drift remains moderate.
As can be seen in Figure 1, 
the concavity of $g_p$ is gradually reduced in time, but does not disappear entirely
in these models.
Although the early evolution of SNRs may not be realistic in our simulations,
this exercise demonstrates that it is important to follow the initial evolution of SNRs, 
including $u_{\rm s}(t)$ and $B(r,t)$, in order to establish the CR spectrum around
$E_{\rm p,max}$.

Figure 2 shows the spatial profile of the CR pressure,
the volume-integrated distribution functions and $G_{\rm p,e}(p)$ 
for protons and electrons at $t/t_{\rm o}=1$,
and the time-integrated spectrum of escaped protons, $\Phi_{\rm esc}(p)$,
at $t/t_{\rm o}=10$ for the same models shown in Figure 1.
Once again, the proton spectrum, $G_{\rm p}$, clearly depends sensitively on 
the magnetic field profile in the precursor, as well as its evolution.
The slope of $G_{\rm p}(p)$ near $p_{\rm p,max}$ approaches $q\sim 3.8$ for the WM1 model and
$q \sim 3$ in the WM2 - WM4 models (see also Figure \ref{fig5}), 
which results from the early nonlinear evolution of DSA.
Hence the $\pi^0$ decay emission in these models could not explain observed $\gamma$-ray 
spectra of some young SNRs, which indicate proton spectra steeper than $F_{\rm p}(p) \propto p^{-4}$.

In the lower left panel of Figure 2, the sum of $G_{\rm p}(p,t)+\Phi_{\rm esc}(p,t)$ is shown in thick lines,
while $\Phi_{\rm esc}(p,t)$ itself is shown in thin lines.
Assuming that the CRs included in $G_{\rm p}$ will be injected eventually into the ISM when the SNR weakens 
and merges to the ambient medium,
this sum at large $t$ would represent the total, time-integrated proton spectrum that 
the model SNR injects into the ISM.
The shape of this integrated spectrum in the four models would be mostly
inconsistent with the Galactic CR proton spectrum around the first knee, because
the spectrum is too flat just below the cutoff.

As demonstrated in the lower right panel of Figure 2, the volume integrated electron spectrum, $G_{\rm e}(p)$, 
steepens approximately by one power of the momentum compared to the
proton spectrum due to radiative cooling above
a break momentum, $p_{\rm e,br}(t)\propto (B_2^2~t)^{-1}$. 
In the WM1 model (solid line), there is an additional peak of $G_{\rm e}(p)$ at a higher energy, 
which is close to the maximum momentum at the subshock, $p_{\rm e,max}$.
This component comes from the electron population in the upstream region, which cools
much less efficiently due to weaker magnetic field there \citep{ekjm11}.
In the other models, magnetic fields are stronger in the precursor, so upstream electrons
have cooled as well.
One can see that in these warm-ISM models the X-ray synchrotron emitting electrons would cut off
at the break energy, $E_{\rm e,br}\sim 0.5-1$ TeV at $t/t_{\rm o}=1$, depending mainly on the strength of $B_2$.

Figure~\ref{fig3} shows for comparison the results of the warm ISM models with downstream Alfv\'enic drift; 
i.e., models 'WM*$_{\rm ad}$'.
Compared to the WM* models shown in Figure 2,
due to the inclusion of downstream Alfv\'enic drift,
the CR acceleration is less efficient, the precursor is weaker, and the shock decelerates less.
As a result, the shock radius is slightly larger in these models, compared to the WM* models.
A weaker precursor also leads to weaker MFA, 
with reduced $B_1\la 100 \muG$ and $B_2\approx 300\muG$ at $t/t_{\rm o}=1$ (see
Figure 6).
In the WM2$_{\rm ad}$, WM3$_{\rm ad}$, and WM4$_{\rm ad}$ models, the high energy end of $F_{\rm p}$ is still much flatter 
than $p^{-4}$, while it is close to $p^{-4}$ in WM1$_{\rm ad}$ model.
Below $p<10^4 m_{\rm p} c$, the CR proton  spectra in the  WM2$_{\rm ad}$, WM3$_{\rm ad}$ and WM4$_{\rm ad}$ models become as steep as $p^{-4.2}$.
Since in those models the proton spectral slopes become as flat as $p^{-3}$
just below their cutoffs, 
the concave curvatures in the spectra become more severe than in the models
without downstream Alfv\'enic drift, WM*.
These models, because of their broader magnetic field precursors, experience
more rapid acceleration early on. Later, as the shocks slow down, and
Alfv\'enic drift becomes more significant, and acceleration of CRs injected
at later times is less efficient.
This effectively leaves an ``island'' of CRs
at the top of the momentum distribution. Once again, this serves as a reminder
that the details of early DSA strongly influence the form of the particle
distribution for a long time.
In these WM*$_{\rm ad}$ models, the electron break energy, $E_{\rm e,br}~\sim 1-3$ TeV at $t/t_{\rm o}=1$. It is slightly higher than that
of the analogous WM* models, because of weaker magnetic fields that result in 
reduced cooling.

We ran two WM1 simulations with increased precursor width due to FEB
placement, $L = \zeta r_{\rm s}$ (WM1$_{\rm feb2,3}$, $\zeta = 0.25, 0.5$),
and one with slow (Bohm) diffusion throughout the precursor (WM1$_{\rm Bohm}$, $K(r)=1$ for $r>r_{\rm s}$).
Figure 4 compares the WM1 model ($\zeta=0.1$),
the WM1$_{\rm feb3}$ model and the WM1$_{\rm Bohm}$ model.
We note the results of WM1$_{\rm feb2}$ are essentially the same as those of WM1$_{\rm feb3}$ model,
and so not shown here.
The diffusion length of protons with $p_{\rm p,max}/m_{\rm p}c=10^5$ is
$l_{\rm max}\approx 0.41$pc for $B_0=5\muG$ and $u_{\rm s}=5000\kms$. 
Since $l_{\rm max}$ is greater than the FEB distance, $L=0.32$pc at $t/t_{\rm o}=1$
in the WM1 model, 
the high energy end of the CR spectrum is strongly affected by particle escape through the FEB.
In the WM1$_{\rm feb2}$ and WM1$_{\rm feb3}$ models, on the other hand, $l_{\rm max}< L=0.8-1.6$pc.
So, escape of the highest particles is not significant, and both WM1$_{\rm feb2,3}$ models have similar CR spectra
extending to $p_{\rm p, max}$ slightly higher than that in WM1 model.
One can see that the overall distribution of $P_{\rm CR}$ is about the same in these
models, except far upstream, near $r_{\rm FEB}$, where $P_{\rm CR}$
is dominated by the highest energy particles that can diffuse to the FEB location.
Referring now to the comparison between the WM1 and WM1$_{\rm Bohm}$ models,
we can see that $G_{\rm p}$ in the WM1$_{\rm Bohm}$ model (dot-dashed line) extends
to higher $p_{\rm p, max}$ by a factor of three or so.
In addition, the spatial profile of $P_{\rm CR}$ is slightly broader, compared to 
the WM1 model.
These differences have similar causes; namely, reduced CR escape at high energies,
reflecting stronger scattering upstream of the subshock in the WM1$_{\rm Bohm}$ model.
Note that the introduction of a FEB upstream of the shock 
or a diffusion scale parameter, $K(r)$, in the precursor does not in fact soften
the CR spectrum near $p_{\rm p,max}$. Instead, it simply affects where an exponential 
cutoff sets in, without altering the CR spectrum just below $p_{\rm p,max}$.

Figure \ref{fig5} compares the volume integrated proton spectrum, $G_{\rm p}$ 
and its slope for different models (at $t/t_{\rm o}=1$ 
for the warm-ISM models and at $t/t_{\rm o}=0.5$ for the hot-ISM models).
Most of these behaviors have already been addressed for the warm-ISM
simulations, but this representation
provides a good general summary and a simple illustration of the comparative properties of the hot-ISM models.
In the first and second rows from the top, we show
WM* models with $u_{\rm w,2}=0$ and WM*$_{\rm ad}$ models with $u_{\rm w,2}=-v_{\rm A,2}$, respectively.
In the third row from the top, the models with the different FEB position are compared:
$\zeta=0.1$ (WM1), 0.25 (WM1$_{\rm feb2}$), and 0.5 (WM$_{\rm feb3}$).
As mentioned previously, we can see here that the WM1$_{\rm feb2}$ (red dotted lines) and WM1$_{\rm feb3}$ (blue dashed) models are almost identical, 
while $p_{\rm p, max}$ of the integrated spectrum is slightly lower in WM1 model (black solid).
For the WM1$_{\rm Bohm}$ model with the Bohm-like diffusion coefficient, $p_{\rm p, max}$ is 
higher than other models because of smaller $\kappa$.

In the bottom row of Figure 5, the three hot-ISM models, HM1, HM1$_{\rm ad}$, and HM1$_{\rm li}$, are illustrated.
In these models
DSA is less efficient compared to the warm-ISM models, because of smaller
sonic Mach number, $M_{\rm s}$,
and because MFA is less efficient in response to smaller Alfv\'enic Mach number, $M_{\rm A,0}$.
As a result, the CR spectra deviate only slightly from the
test-particle power-law and the downstream magnetic fields are weaker.
We also show in the bottom row of Figure \ref{fig5} results from the one case we computed with a
reduced CR injection rate; model HM1$_{\rm li}$.
As expected, because $P_{\rm CR}$ is reduced, the precursor flow is less modified
and the CR acceleration is almost in the test-particle regime.

From the results illustrated in Figure 5 it seems difficult for the typical SNR parameters 
considered here to obtain a proton momentum spectrum steeper
than $p^{-4}$ near $p_{\rm p,max}$ even when the nominally rather strong
influences of postshock Alfv\'enic drift are included.
The comparison of different models in Figure 5 further illustrates how the spectrum 
of accelerated
CRs depends on the nonlinear interplay among MFA, Alfv\'enic drift and particle escape.

Turning briefly to the simulated CR electron properties, since they are responsible for
the radio through X-ray nonthermal emissions in SNRs, we note
for $p<p_{\rm e,br}$ that the slopes of $G_{\rm e}$ and $G_{\rm p}$ should be similar, because the electron
spectrum is not affected by radiative cooling at those momenta.
According to Figure 5, the slope of $G_{\rm e}(p)$ 
near $p/m_{\rm p}c\sim 1$ ($\gamma_{\rm e} \sim 2000$) would be about $4.2$, 
virtually independent of the  model parameters.
So the radio spectral index is expected to be similar in these models.

Figure 6 shows the time evolution of the various dynamical shock properties for different models,
including the density compression factors, amplified magnetic field strengths,
the postshock CR pressure, CR injection fraction and the fraction of the explosion
energy transferred to CRs
for the models without (left column) or with (right column) postshock
Alfv\'enic drift.
The left column also includes for comparison the WM1$_{\rm li}$ model with a lower injection.
Several of these comparisons have already been addressed. 
According to Equation (\ref{Bpre}), the MFA factor, $B_1/B_0$, depends on the precursor
strength (i.e., $U_1$) and the Alfv\'enic Mach number, $M_{A,0}$.
So the preshock magnetic field, $B_1$, increases rapidly in the early stage during which
the precursor develops, but later it decreases in time with diminishing $M_{A,0}$.
In the descending order of the effective width of the magnetic field precursor, 
{\bf{M2}, \bf{M4}, \bf{M3}, \bf{M1}},
the precursor grows and the value of $B_1(t)$ peaks at progressively earlier times.
After reaching its peak value, the $B_1$ values decline as $B_1 \approx 136 \muG (t/t_o)^{-3/5}$
and become similar in all four models.
Here the straight dotted lines show the limiting magnetic field, $B_{\rm sat} = 168\muG (t/t_o)^{-3/5}$,
given in Equation (\ref{Bsat}).
One can see that $B_1$ stays below $B_{\rm sat}$ for the models considered here.

From the third row of Figure 6 we can see that the MFA model profile strongly
influences the early evolution of the postshock CR pressure, $P_{\rm CR,2}$. 
But once the precursor growth and MFA saturate, the postshock CR pressure is
similar for a given model class (postshock Alfv\'enic drift on or off).
Because of the steepening of the CR spectrum (see Figure 5), 
$P_{\rm CR,2}$ is smaller in WM*$_{\rm ad}$ models, compared to WM* models. 
But the difference is less than a factor of two.
The bottom panels of Figure \ref{fig6} show that by the end of the simulations 
about 30\% of the SN explosion energy is transferred to CRs in WM* models,
while WM*$_{\rm ad}$ models are somewhat less efficient, as expected. 
As can be seen in Figures 2 and 3, $P_{\rm CR,2}$ is smaller but the shock radius $r_s$ is larger
in WM*$_{\rm ad}$ models, 
resulting in a rather small difference in the volume integrated energy, $E_{\rm CR}$,
between the two model classes.
A reduced CR injection rate also reduces
this energy transfer in the WM1$_{\rm li}$ model.
Thus, in our models the amount of the CR energy produced in the SNRs is smaller than what had been
reported in some previous DSA simulations in which the Alfv\'enic drift was not included
or Alfv\'en speed in the background field, $B_0$, was adopted \citep[e.g.][]{bv97, bkv09, kj06}.
Even though the CR acceleration efficiency is reduced in these simulations, 
the estimated values
are still sufficient to replenish CR energy escape from the Galaxy.

\subsection{Emissions}

It is useful to know how the differences in CR particle spectra are translated 
into differences in nonthermal emissions spectra \citep[e.g.,][]{ekjm11}.
For a power-law electron distribution, $f_{\rm e}(p)\propto p^{-q_{\rm e}}$, the radiation spectra
due to synchrotron and iC processes will have a power-law form, $F_{\nu}\propto
\nu^{-\alpha}$ with a slope $\alpha_{\rm syn}=\alpha_{\rm iC}=(q_{\rm e}-3)/2$. 
For the test-particle slope, $q_{\rm e}=4$, $\alpha_{\rm syn}=\alpha_{\rm iC}=0.5$, 
so $\nu L_{\nu}\propto \nu^{+0.5}$, where $L_{\nu}$ is the 
luminosity spectrum in units of $\rm erg~ s^{-1} eV^{-1}$.
The same power-law momentum distribution for protons, $f_{\rm p}(p)\propto p^{-q_{\rm p}}$ 
with a cutoff at $E_{\rm p,max}$,
$\pi^0$ decay emission spectrum has roughly a power-law form with $\alpha_{\pi^0}=q_{\rm p}-3$,
for the photon energy $80 {\rm MeV}\la E_{\gamma}\la 0.1 E_{\rm p,max}$ \citep{kej12}.
So for $\pi^0$ decay emissions with $q_{\rm p}=4$, $\nu L_{\nu}$ has a flat shape 
($\alpha_{\pi^0}+1=0$) typically for the 100 MeV - 10 TeV band
(note: $\nu=10^{24}$ Hz corresponds to $E_{\gamma}=4$ GeV).

In Figure 7 the volume integrated radiation spectra, $\nu L_{\nu}$, are shown for different
models, where $K_{\rm e/p}=10^{-4}$ is adopted for convenience.
In the upper two panels the contributions from the different processes are compared
for WM1$_{\rm ad}$ and W4 models.
The lower left panel shows the results for four models with different profiles of $B(r)$
in the precursor: WM1, WM2, WM3, and WM4 models.
The results for the hot-ISM models are shown in the lower right panel.

As shown in Figure 5, the slope of the electron spectrum is almost universally
$q_{\rm e}\approx 4.2$ for $\gamma_{\rm e} \sim 10^3-10^5$, 
so the spectral shape of the radio synchrotron emission should be similar 
in different models.
On the other hand, the X-ray synchrotron emission can be affected by the radiative cooling.
In the hot ISM models, the postshock CR electron population has not cooled significantly
by $t/t_{\rm o}=0.5$, so $\nu L_{\nu}$ has a relatively sharp peak in the X-ray band.  
In the warm ISM models with much stronger amplified magnetic fields,
the rise of $\nu L_{\nu}$ in X-ray synchrotron emission before the cutoff is 
flattened by the radiative cooling.
This demonstrates that the spectral shape in the X-ray band is affected by the
evolution of $B(r,t)$ as well as the shock dynamics, $u_{\rm s}(t)$.
In particular, for the WM4 model a distinctive signature of electron cooling can be seen 
in iC emission for $E_{\gamma}=1-100$ GeV 
(in the case where iC emission dominates over $\pi^0$ decay emission).

With the assumed value of $K_{\rm e/p}$
and a background radiation field representative of the galactic plane,
$\pi^0$ decay emission dominates
over electronic iC scattered emission in the GeV-TeV $\gamma$-ray band.
As can be seen in Figures 2 and 3, the high energy end of CR proton spectra
vary widely among different models, depending on the models for MFA and 
Alfv\'enic drift, so the ensuing $\pi^0$ decay emission spectra can be very different.
Model WM1$_{\rm ad}$ shown in the top left panel has the proton cutoff 
at $E_{\rm p,max}\sim 10^4$ GeV and the $\gamma$-ray cutoff at $E_{\gamma}\sim 4$ TeV 
($10^{27}$ Hz). For this model, the shape of $\nu L_{\nu}$ due to $\pi^0$ decay
is slightly steeper than the flat spectrum expected for the test-particle power law with $q_{\rm p}=4$.
The warm ISM models with other MFA models, {\bf M2-M4}, have the $\gamma$-ray cutoff at
higher energies, $E_{\gamma}\sim 10-100$ TeV,
and their $\pi^0$ decay emission spectra show a concave curvature.
Such a signature of nonlinear DSA has not been observed in real SNRs,
so these models are probably unrealistic and a significant revision for
the current DSA theory is called for.
But as stated before, our focus here is to demonstrate how
different models for MFA, Afv\'enic drift and FEB affect
the high energy end of the CR proton spectra and their nonthermal emission spectra.
We note that the sharp feature near $E_{\gamma}\sim 100$ MeV comes from 
the low energy cross-section for the $p-p$ collision.

\section{SUMMARY}

The most direct evidence for the particle acceleration at SNR shocks can be provided 
by multi-wavelength observations of nonthermal radiation emitted by CR protons and electrons
\citep[e.g.][]{bkv12,morlino12,kej12}.
Recent observations of young SNRs in the GeV-TeV bands seem to imply that the accelerated proton 
spectrum might be much steeper than predicted by conventional nonlinear DSA theory, which is 
based on some simplifying assumptions for wave-particle interactions 
\citep[e.g.][]{acero10,acciari11,giordano12}.
Thus a detailed understanding of plasma physical processes at collisonless shocks
is essential in testing the DSA hypothesis for the origins of Galactic CRs 
\citep[e.g.][]{capri12,kang13}.
In this study, we have explored how magnetic field amplification, drift of scattering centers,
and particle escape could affect the outcomes of nonlinear DSA at the outer shock of Type Ia SNRs,
implementing some phenomenological models for such processes into the exiting DSA simulation code.

Given the current lack of full understanding of different essential processes,
we have considered several heuristic models and included a moderately
large number of models and parameters in order to examine underlying
model sensitivities. We do not claim that any of these models accurately
represent real, young SNRs. We have, in particular, considered: 
{\bf a)} four different models for  magnetic field amplification (MFA), $B(r,t)$, in the precursor,
{\bf b)} inclusion of wave damping ($\rightarrow$ plasma heating) through a parameter, $\omega_{\rm H}$,
{\bf c)} Alfv\'enic drift, with allowance for super-gyro-scale field disorder,
through a drift speed adjustment parameter, $f_{\rm A}$, {\bf d)} inclusion of
downstream, postshock Alfven\'ic drift in some cases, {\bf e)} a free escape
boundary (FEB) with adjustable scale through a parameter, $\zeta$, {\bf f)} 
reduced CR scattering efficiency towards the front of the shock
precursor, through a diffusion scale parameter, $K(r)$, and {\bf g)} 
variation of the thermal leakage injection rate through a  parameter, $\epsilon_B$.
In these, kinetic DSA simulations the time-dependent evolution of
the pitch-angle averaged phase-space distribution functions of CR protons and electrons
are followed along with the coupled, dynamical evolution of an initially Sedov-Taylor blast wave.
Radiation spectra from the CR electrons and protons have been calculated 
through post processing of the DSA simulation data.

Since the spatial profile of the amplified magnetic field, $B(r,t)$, 
is not in general a simple step function,
the time-dependent evolution of the particle acceleration depends on
interplay between smaller diffusion coefficient and faster Alfv\'enic drift, 
which have opposite effects on DSA.
Stronger magnetic fields result in smaller scattering lengths and faster acceleration,
leading to higher $p_{\rm max}$ and greater precursor compression.
On the other hand, faster Alfv\'enic drift away from the shock 
steepens the CR spectrum and reduces the CR acceleration efficiency
\citep{pzs10,capri12}.

The main results can be summarized as follows.

1. The high energy end of the proton spectrum depends sensitively on the strength and
profile of the amplified magnetic field in the precursor.
For typical SNR shock properties considered here, the maximum proton energy can reach up to 
$E_{\rm p,max}\approx 1$ PeV except for one MFA model ({\bf M1}, Equation (\ref{Bpre})) 
which has a very
thin region of amplified magnetic field in the precursor, such that
its effective width is less than the diffusion length of the highest
energy particles produced in the other MFA models. 
Consequently, in the {\bf M1} models, in which the magnetic field is not amplified
significantly for most of the precursor,
DSA is too slow to reach such high energies.
This model exhibits relatively less CR spectral concavity as a byproduct of
its relative inability to accelerate to very high energies.

2. The MFA models {\bf M2} - {\bf M4} given in Equations (\ref{Bpre2})-(\ref{Bpre4})
produce broader magnetic field precursors than the {\bf M1} model. In those models
the CR spectrum is indeed steepened significantly for $E<10$ TeV
by fast Alfv\'enic drift.
But the CR spectrum for $E>10$ TeV shows a strong concave curvature, since the high energy
ends of the spectra are established early on, when magnetic fields are being amplified through
the initial development of the precursor. In fact, the CR spectra just below
$E_{\rm p,max}$ in these models can approach the very hard spectral slope, $E^{-1}$.

3. In the models with postshock Alfv\'enic drift ($u_{\rm w,2} \approx - v_{\rm A}$),
the CR spectrum is slightly softer than that of the models without it, 
although it remains somewhat concave.
In these models, the CR acceleration efficiency is reduced by a factor of less than two
and MFA in the precursor becomes also weaker, compared to the models without postshock Alfv\'enic drift.

4. Reduced CR scattering efficiencies far upstream of the shock due, for
instance to the turbulent field being dominated by small scale, non-resonant 
fluctuations, and resulting escape of the highest energy
particles also regulate the high energy end of the CR proton spectrum in important
ways (i.e., $K(r)\ge1$).
We demonstrated, in addition, that the CR proton spectrum near the high energy cutoff is strongly
influenced by the FEB location if the high energy diffusion length, $l_{\rm max}$,
approaches the width of the precursor; i.e., if $L<l_{\rm max}$,
where $L=\zeta r_{\rm s}$ is the FEB distance.
However, the introduction of a diffusion scale parameter, $K(r)$, or a FEB upstream of
the shock simply lowers $p_{\rm p,max}$ where the exponential cutoff sets in, 
rather than steeping the CR spectrum $p\la p_{\rm p,max}$.

5. Nonthermal radiation from SNRs carries significant information about nonlinear DSA
beyond the simple test-particle predictions. In particular, the shape of X-ray
synchrotron emission near the cutoff is determined by the evolution of the amplified
magnetic field strength as well as the shock dynamics.

6. Since the high energy end of the CR proton spectrum consists of the particles
that are injected during the early stages of SNRs, the spectral shape of $\pi^0$ decay
emission near the high energy cutoff depends on the time-dependent evolution of
the CR injection, MFA, and particle escape
as well as the early dynamical evolution of the SNR shock.
These properties, in return, depend on dynamical feedback from the CRs and
MFA. The end results are not likely well modeled by a succession of independent,
static solutions.

This study demonstrates that a detailed understanding of plasma physical processes
operating at collisionless shocks is crucial in predicting the CR energy spectra accelerated
at SNR shocks and nonthermal emissions due to those CRs.

\acknowledgements

H.K. was supported by Basic Science Research Program through
the National Research Foundation of Korea (NRF) funded by the Ministry
of Education, Science and Technology (2012-001065).
T.W.J. was supported in this work at the University of Minnesota
by NASA grant NNX09AH78G, NSF grant AST-1211595
and by the Minnesota Supercomputing Institute for Advanced Computational 
Research.
P.P.E. was supported by Harvard Research Computing and the Institute for Theory and Computation at the Center for Astrophysics.
The authors would like to thank the anonymous referee for the 
constructive suggestions and comments.
H.K. also thanks Vahe Petrosian and KIPAC for their
hospitality during the sabbatical leave at Stanford University
where a part of the paper was written.

\clearpage

\begin{deluxetable} {lrrrrrrrr}
\tablecaption{Model Parameters$^{\rm a}$}
\tablehead{
\colhead {Name$^{\rm b}$} & \colhead{$n_{\rm H}$} & \colhead{$T_0$} & \colhead{$B(r)$} & 
\colhead{ $f_{\rm A}~ ^{\rm c}$}& \colhead{$\omega_{\rm H}~ ^{\rm }$}& \colhead{~$\zeta~ ^{\rm d}$}&\colhead{$u_{\rm w,2}~^{\rm }$}& \colhead{$\epsilon_B$} \\
\colhead {} & \colhead {$(\rm cm^{-3})$} & \colhead {(K)} & \colhead { for $r>r_{\rm s}$} & \colhead {} &
\colhead {} & \colhead {}  & \colhead {} & \colhead {}  }

\startdata
WM1 & 0.3   & $3\times10^4$ & Eq. (2)  & 0.5 &0.5 & 0.1 & 0 & 0.215 \\
WM1$_{\rm ad}$ & 0.3   & $3\times10^4$ & Eq. (2)  & 0.5 &0.5 & 0.1 & $-v_{\rm A}$ & 0.215 \\
WM1$_{\rm li}$ & 0.3   & $3\times10^4$ & Eq. (2) & 0.5 &0.5 & 0.1 & 0 & 0.2  \\
WM1$_{\rm Bohm}$$^{\rm e}$  & 0.3   & $3\times10^4$ & Eq. (2) & 0.5 &0.5 & 0.1 & 0 & 0.215  \\
\\
WM1$_{\rm feb2}$ & 0.3   & $3\times10^4$ & Eq. (2)  & 0.5 &0.5 & 0.25 & 0 & 0.215 \\
WM1$_{\rm feb3}$ & 0.3   & $3\times10^4$ & Eq. (2)  & 0.5 &0.5 & 0.5 & 0 & 0.215\\
\\
WM2 & 0.3   & $3\times10^4$ & Eq. (3)  & 0.5 &0.5 & 0.1 & 0 & 0.215 \\
WM2$_{\rm ad}$ & 0.3   & $3\times10^4$ & Eq. (3)  & 0.5 &0.5 & 0.1 & $-v_{\rm A}$ & 0.215\\
\\
WM3 & 0.3   & $3\times10^4$ & Eq. (4)  & 0.5 &0.5 & 0.1 & 0 & 0.215 \\
WM3$_{\rm ad}$ & 0.3   & $3\times10^4$ & Eq. (4)  & 0.5 &0.5 & 0.1 & $-v_{\rm A}$ & 0.215 \\
\\
WM4 & 0.3   & $3\times10^4$ & Eq. (5)  & 0.5 &0.5 & 0.1 & 0 & 0.215 \\
WM4$_{\rm ad}$ & 0.3   & $3\times10^4$ & Eq. (5)  & 0.5 &0.5 & 0.1 & $-v_{\rm A}$ & 0.215 \\
\\
HM1  & 0.01 &$10^6$ & Eq. (2) & 0.5  &0.5  & 0.1 & 0 & 0.215\\
HM1$_{\rm ad}$  & 0.01 &$10^6$ & Eq. (2) & 0.5  &0.5  & 0.1 & $-v_{\rm A}$ & 0.215\\
HM1$_{\rm li}$  & 0.01 &$10^6$ & Eq. (2) & 0.5  &0.5  & 0.1 & 0 & 0.2\\

\enddata
\tablenotetext{a}{For all the models, the SN explosion energy is $E_0=10^{51} {\rm ergs}$, and $B_0=5\mu$G. 
The normalization constants are $r_{\rm o}=3.18$pc and $t_{\rm o}=255$ years for the warm ISM models
and $r_{\rm o}=9.89$pc and $t_{\rm o}=792$ years for the hot ISM models. All simulations
are initialized from ST conditions at $t/t_0 = 0.2$.}
\tablenotetext{b}{`W' and `H' stand for the warm and hot phase of the ISM, respectively. 
'M1-M4' indicates the different MFA models, $B(r)$, in the precursor, as defined
in section 2.2.}
\tablenotetext{c}{See Equation (\ref{vA}) for the Alfv\'en drift parameter, $f_{\rm A}$. }
\tablenotetext{d}{Free escape boundary is at $r_{\rm FEB}= (1+\zeta) r_{\rm s}(t)$. }
\tablenotetext{e}{Same as WM1 except that $K(r)=1$ (Bohm diffusion) for $r>r_{\rm s}$. }
\
\end{deluxetable}

\begin{figure}
\vspace{+1cm}
\begin{center}
\includegraphics[scale=0.85]{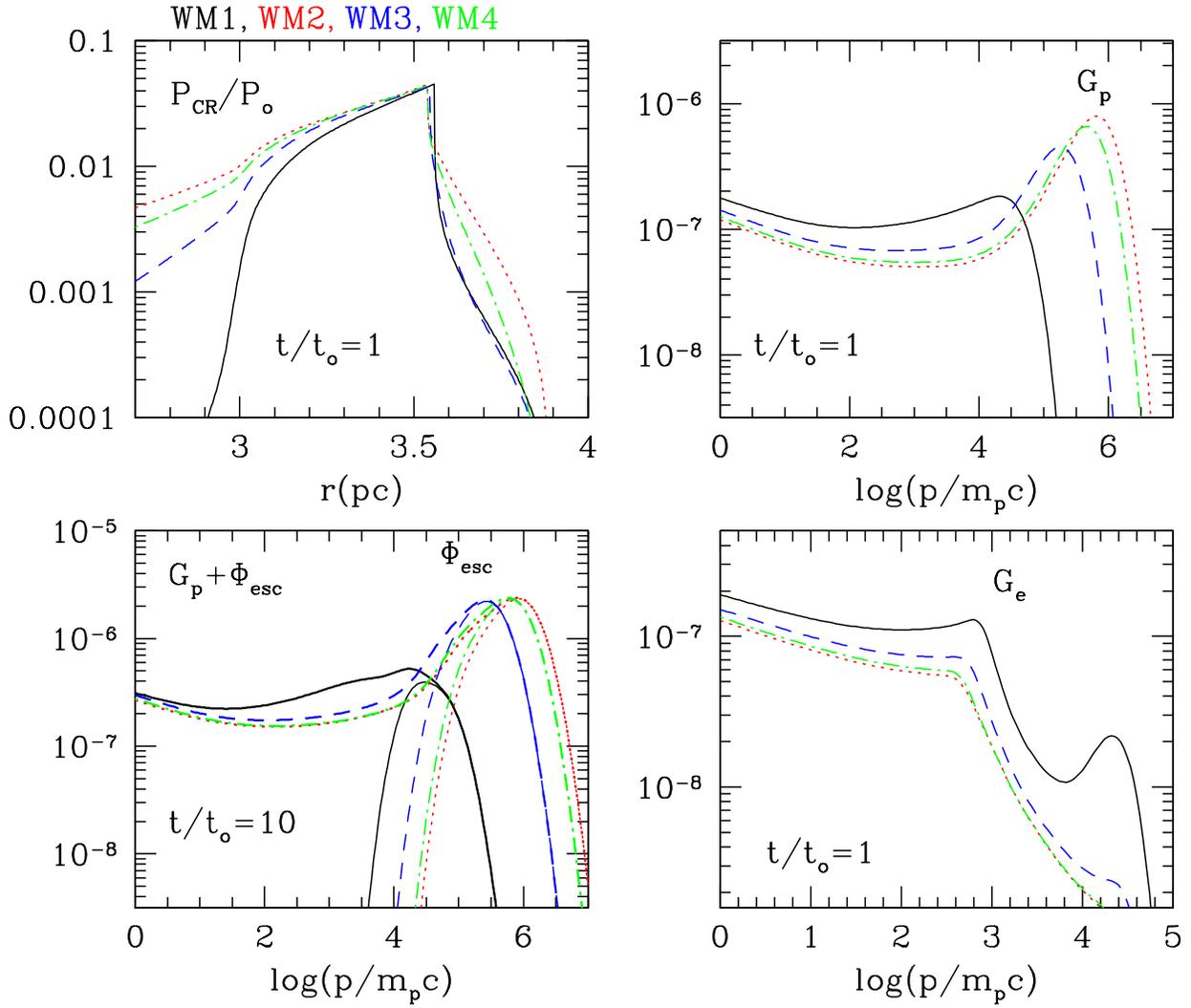}
\end{center}
\vspace{-0.0cm}
\caption{
CR pressure profile and the CR spectra
are shown at $t/t_{\rm o} =1$ ($t_{\rm o}=255$ years) for WM1 (black solid lines), WM2 (red dotted), 
WM3 (blue dashed]) and WM4 (green dot-dashed) models.
The volume integrated distribution functions, $G_{\rm p}(p)$ for protons
and $G_{\rm e}(p)$ for electrons, and the time-integrated spectrum of escaped protons,
$\Phi_{\rm esc}(p)$, are given in arbitrary units. 
In the lower left panel, 
$\Phi_{\rm esc}(p)$ (thin lines), and the sum of $G_{\rm p}(p)$ and $\Phi_{\rm esc}(p)$
(thick lines) are shown at $t/t_{\rm o} =10$ for the same four models.
Here $K_{\rm e/p}=1$ is assumed.
}
\label{fig2}
\end{figure}
\begin{figure}
\vspace{+1cm}
\begin{center}
\includegraphics[scale=0.85]{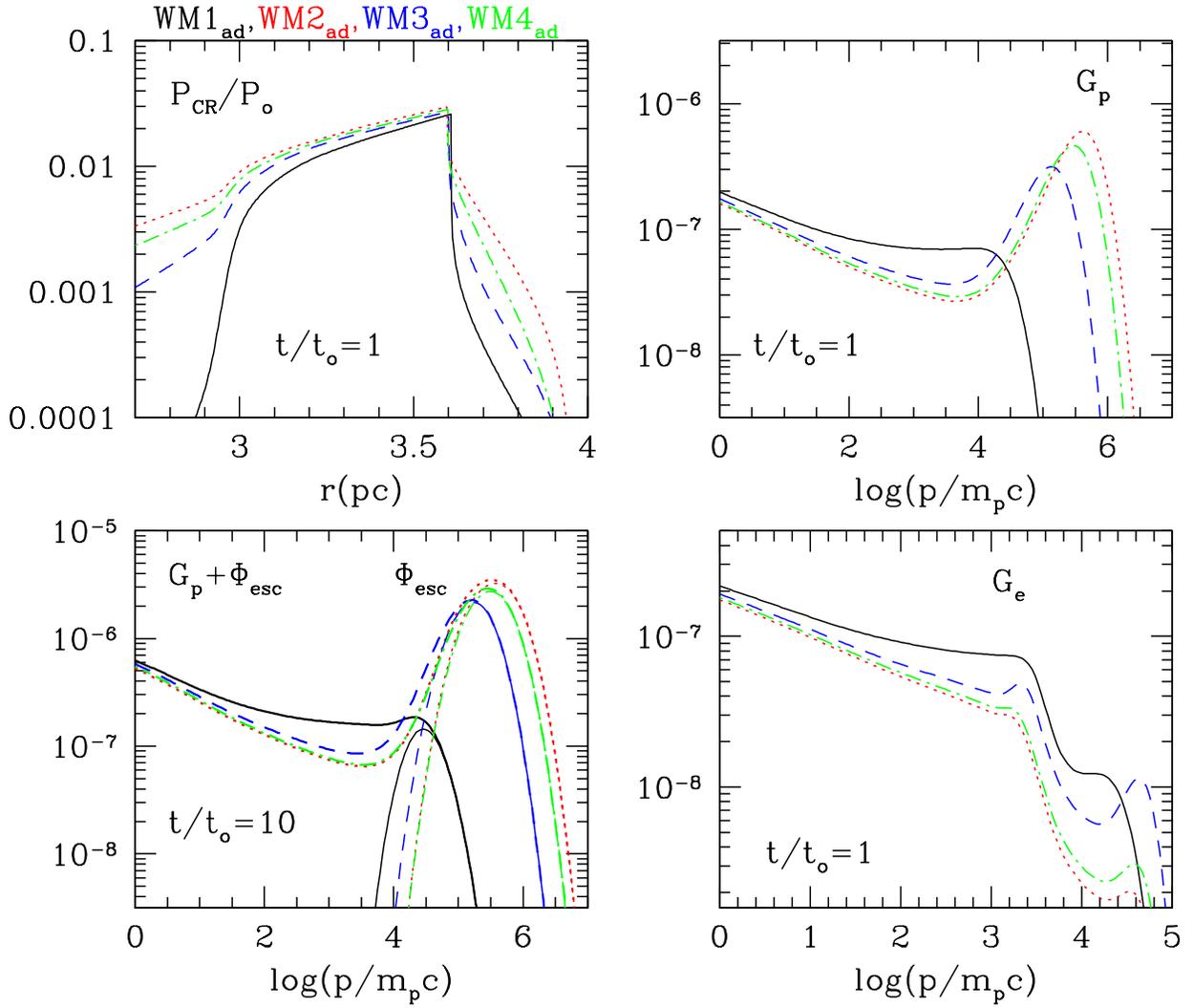}
\end{center}
\vspace{-0.0cm}
\caption{
Same as Figure 2 except the models with downstream Alfv\'enic drift,
WM1$_{\rm ad}$ (black solid lines), WM2$_{\rm ad}$ (red dotted), WM3$_{\rm ad}$ (blue dashed) and WM4$_{\rm ad}$ (green dot-dashed)
models are shown.
}
\label{fig3}
\end{figure}
\clearpage
\begin{figure}
\vspace{+1cm}
\begin{center}
\includegraphics[scale=0.85]{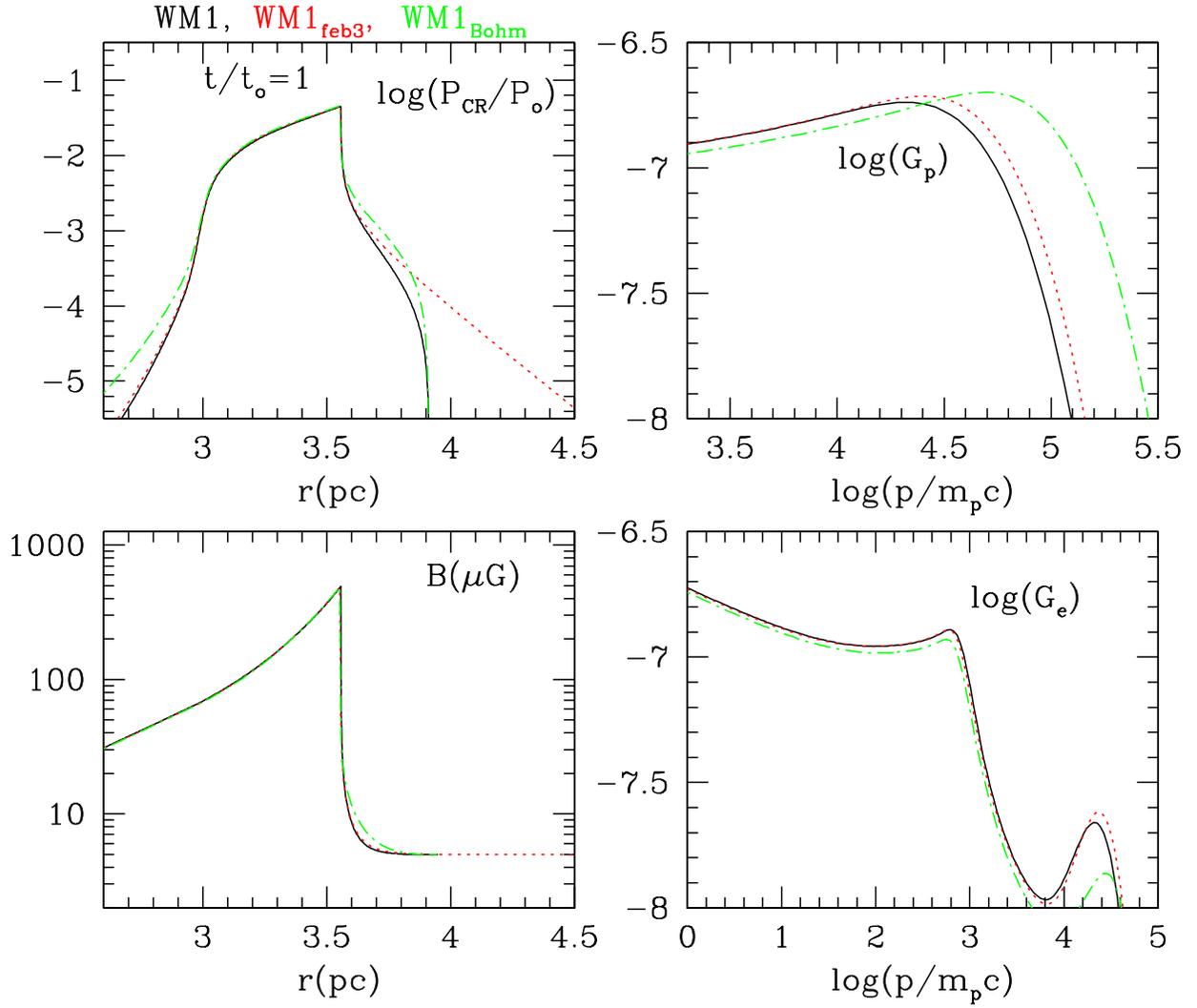}
\end{center}
\vspace{-0.0cm}
\caption{
Profiles of the volume-integrated CR pressure and the magnetic field, and the CR spectra
are shown at $t/t_{\rm o} =1$ for WM1 (black solid lines), WM1$_{\rm feb3}$ (red dotted), 
and WM1$_{\rm Bohm}$) (green dot-dashed) models.
In WM1$_{\rm feb3}$ model the FEB is located at $r=1.5\cdot r_{\rm s}$, whereas
the function $K(r)=1.0$ for $r>r_{\rm s}$ in WM1$_{\rm Bohm}$ model.
Here $K_{\rm e/p}=1$ is assumed.
See Table 1 for model parameters.
}
\label{fig4}
\end{figure}
\clearpage
\begin{figure}
\vspace{+1cm}
\begin{center}
\includegraphics[scale=0.85]{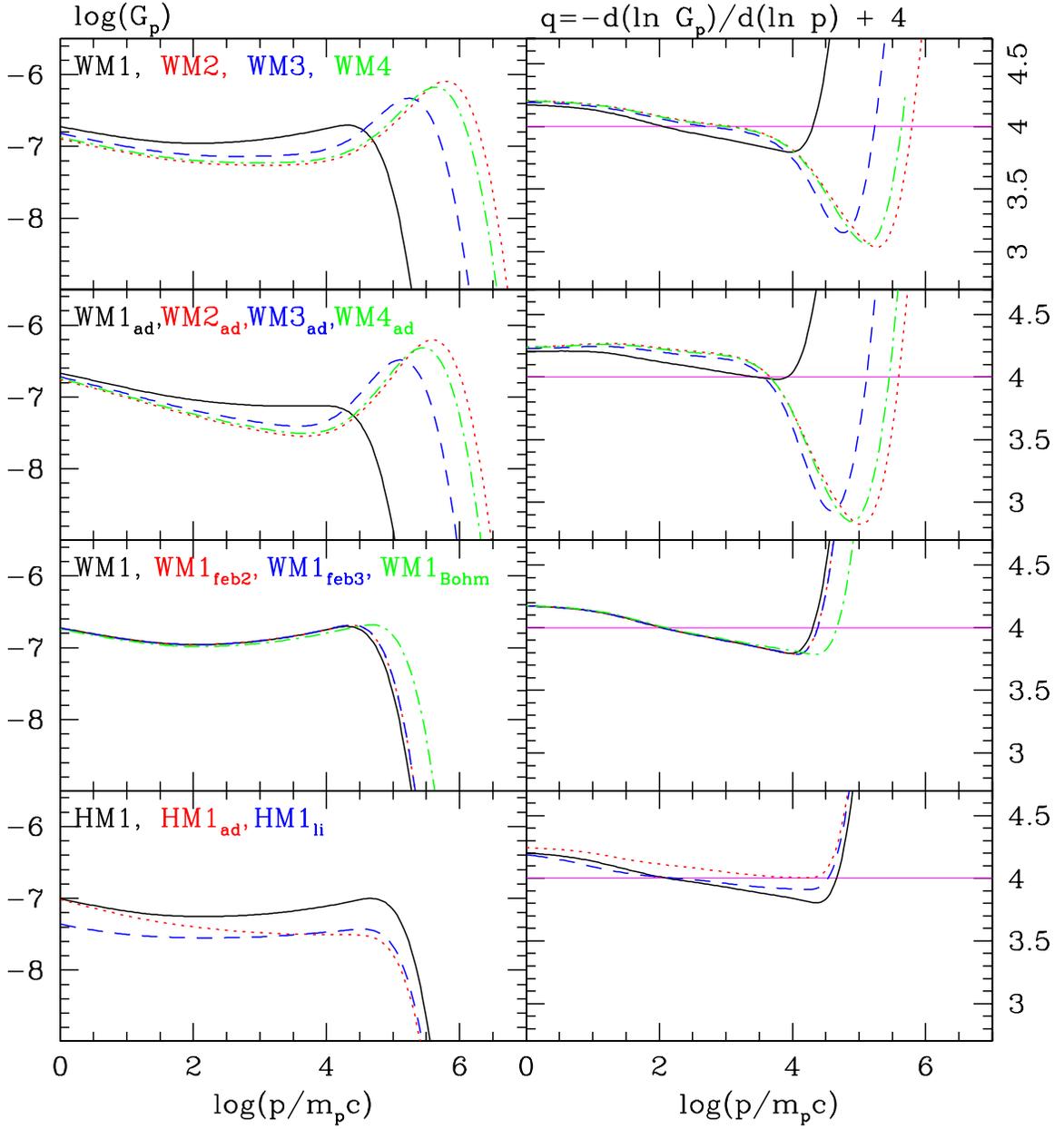}
\end{center}
\vspace{-0.0cm}
\caption{
Left: Volume integrated proton spectrum at $t/t_{\rm o}=1$ for the warm-ISM models (top three rows)
and at $t/t_{\rm o}=0.5$ for the hot-ISM models (bottom row).
Right: 
Power-law slope of $G_{\rm p}$, $ q=-\partial \ln G_{\rm p}/ \partial \ln p + 4$, which are shown in
the left panels.
The horizontal line marks $q=4$.
From top to bottom the line types are chosen as follows: 
WM1 (black solid), WM2 (red dotted), WM3 (blue dashed), WM4 (green dot-dashed);
WM1$_{\rm ad}$ (black solid), WM2$_{\rm ad}$ (red dotted), WM3$_{\rm ad}$ (blue dashed), WM4$_{\rm ad}$ (green dot-dashed);
WM1 (black solid), WM1$_{\rm feb2}$ (red dotted), WM1$_{\rm feb3}$ (blue dashed), 
WM1$_{\rm Bohm}$ (green dot-dashed);
HM1 (black solid), HM1$_{\rm ad}$ (red dotted), HM1$_{\rm li}$ (blue dashed).
}
\label{fig5}
\end{figure}
\clearpage
\begin{figure}
\vspace{+1cm}
\begin{center}
\includegraphics[scale=0.85]{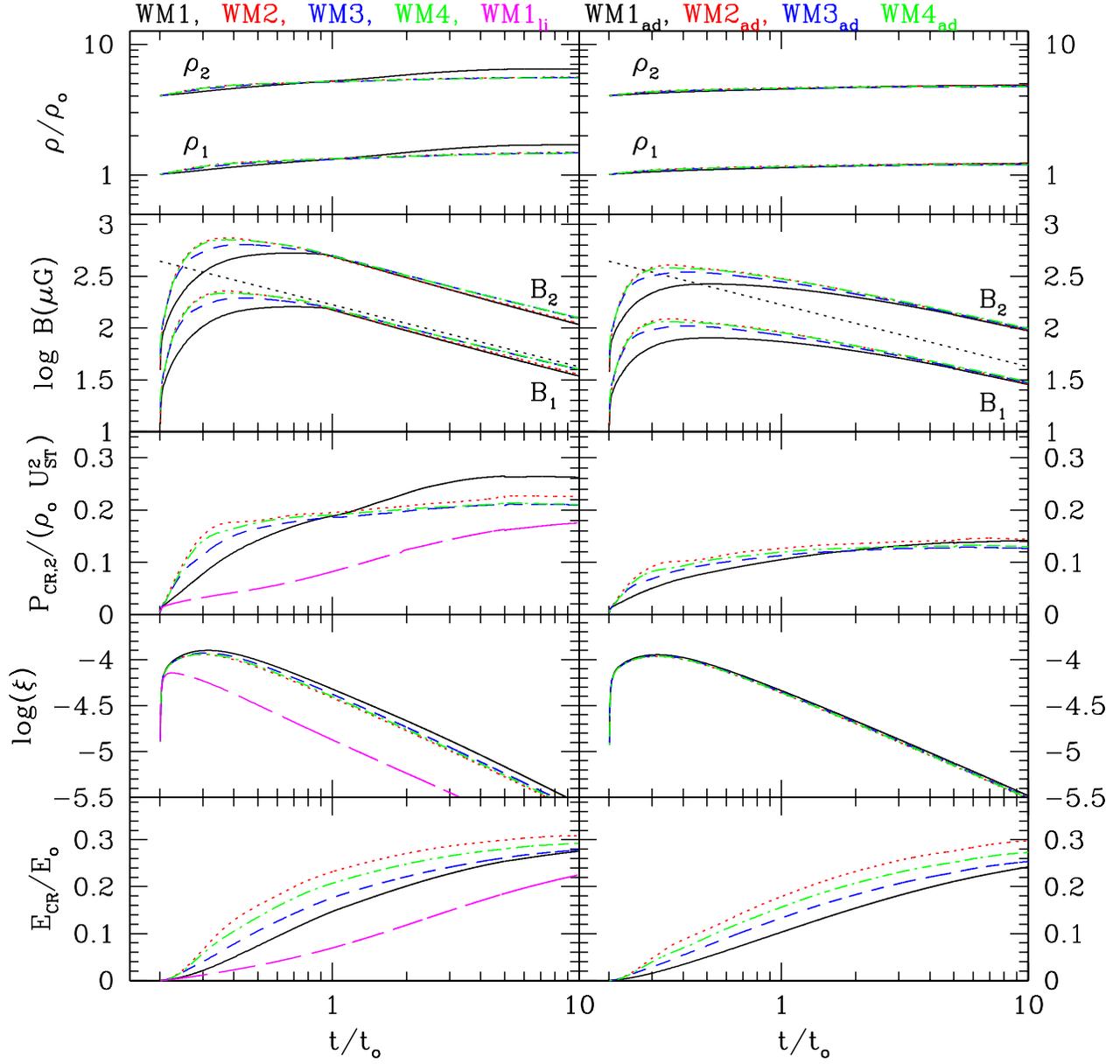}
\end{center}
\vspace{-0.0cm}
\caption{
Time evolution of the gas density $\rho_1$ ($\rho_2$) immediate upstream (downstream) of the subshock,
and magnetic field strengths, $B_1$ and $B_2$, 
postshock CR pressure, $P_{\rm CR,2}$ in units of the ram pressure of the unmodified Sedov-Taylor solution ($U_{\rm ST}/u_{\rm o}=0.46(t/t_{\rm o})^{-3/5}$),
the injection fraction, $\xi$, and the total volume integrated CR energy in units of $E_{\rm o}$ in different models:
WM1 (black solid lines), WM2 (red dotted), WM3 (blue dashed), WM4 (green dot-dashed),
WM1$_{\rm li}$ (magenta long-dashed) in the left column,
and WM1$_{\rm ad}$ (black solid lines), WM2$_{\rm ad}$ (red dotted), WM3$_{\rm ad}$ 
(blue dashed), WM4$_{\rm ad}$ (green dot-dashed) in the right column.
For WM1$_{\rm li}$ model only $P_{\rm CR,2}$, $\xi$, and $E_{\rm CR}$ are shown.
In the second row from the top, the straight dotted lines represent $B_{\rm sat}=(8\pi\cdot 0.005 \rho_0 U_{\rm ST})^{1/2}$.
Note $t_{\rm o}=255\yrs$ for the warm ISM models.
}
\label{fig6}
\end{figure}
\clearpage

\begin{figure}
\vspace{+1cm}
\begin{center}
\includegraphics[scale=0.85]{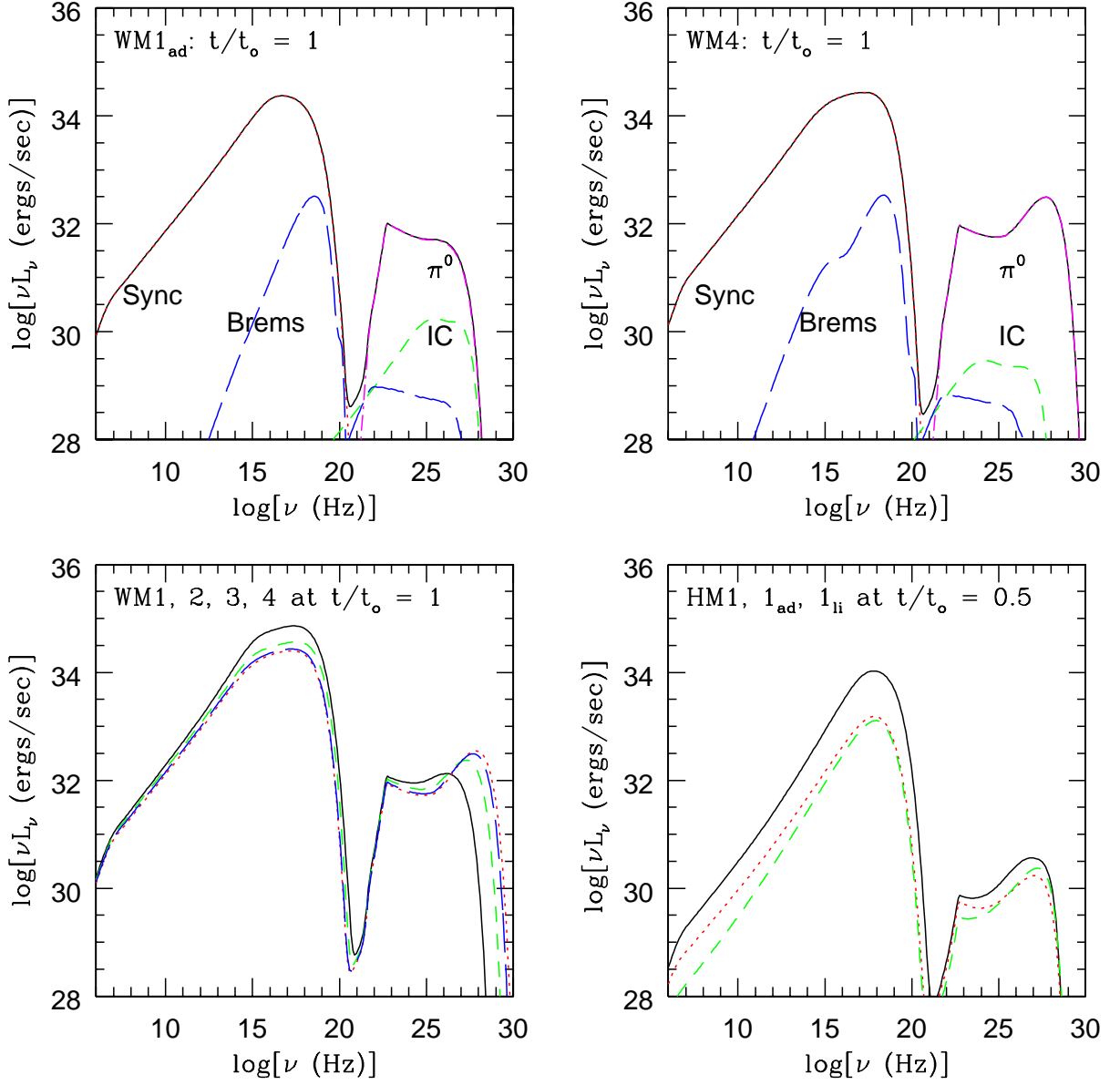}
\end{center}
\vspace{-0.0cm}
\caption{
Volume integrated radiation spectra, $\nu L_{\nu}$ (erg/s), are shown.
Here $K_{\rm e/p}=10^{-4}$ is assumed.
Top left and right panels: synchrotron (red dotted line), bremsstrahlung (blue long dashed), iC scattering 
(green dashed), $\pi^0$ decay (magenta dot-dashed), and the total spectrum (black solid) are
shown at $t/t_{\rm o}=1$ for WM1$_{\rm ad}$ (left) and WM4 (right) models.
Bottom left: Total radiation spectrum at $t/t_{\rm o}=1$ for WM1 (black solid line), 
WM2 (red dotted), WM3 (green dashed), and WM4 (blue long dashed) models.
Bottom right: Total radiation spectrum at $t/t_{\rm o}=0.5$ for HM1 (black solid line), 
HM1$_{\rm ad}$ (red dotted), HM1$_{\rm li}$ (green dashed) models.
See Table 1 for different model parameters.
}
\label{fig7}
\end{figure}
\clearpage

\end{document}